\documentclass[preprintnumbers,amsmath,amssymb,prd]{revtex4}

\usepackage{graphicx}% Include figure files
\usepackage{dcolumn}% Align table columns on decimal point
\usepackage{bm}% bold math
\usepackage{subfigure}

\newcommand{\ihmpc}{\, h\, {\rm Mpc}^{-1}}

\newcommand{\lyaf}{Ly$\alpha$ forest}

\newcommand{\vq}{\mathbf{q}}
\newcommand{\vx}{\mathbf{x}}
\newcommand{\vk}{\mathbf{k}}
\newcommand{\tdtwo}{\tilde{b}_{\delta^2}}
\newcommand{\tstwo}{\tilde{b}_{s^2}}
\newcommand{\tbthree}{\tilde{b}_3}
\newcommand{\tbR}{\tilde{b}_R}
\newcommand{\tadtwo}{\tilde{a}_{\delta^2}}
\newcommand{\tastwo}{\tilde{a}_{s^2}}
\newcommand{\tabthree}{\tilde{a}_3}
\newcommand{\vnabla}{\mathbf{\nabla}}
\newcommand{\ttheta}{\eta}
\newcommand{\tttheta}{\psi}
\newcommand{\vv}{\mathbf{v}}
\newcommand{\fnl}{f_{\rm NL}}

\newcommand{\ssqfac}{S}
\newcommand{\FtwoS}{{F_S^{\left(2\right)}}}
\newcommand{\FthreeS}{{F_S^{\left(3\right)}}}

\begin{document}

\title{Clustering of dark matter tracers: generalizing bias for the coming era 
of precision LSS}

\author{Patrick McDonald}
\email{pmcdonal@cita.utoronto.ca}
\affiliation{Canadian Institute for Theoretical Astrophysics, University of
Toronto, Toronto, ON M5S 3H8, Canada}
\author{Arabindo Roy}
\email{roy@lepus.astro.utoronto.ca}
\affiliation{Department of Astronomy and Astrophysics, University of
Toronto, Toronto, ON M5S 3H8, Canada}

\date{\today}

\begin{abstract}

On very large scales, density fluctuations in the Universe are small, 
suggesting a perturbative model for large-scale clustering of galaxies (or 
other dark matter tracers), in which the galaxy density is written as a Taylor 
series in the local mass density, $\delta$, with the unknown coefficients in 
the series treated as free ``bias'' parameters. We extend this model to include
dependence of the galaxy density on the local values of $\nabla_i\nabla_j\phi$ 
and $\nabla_i v_j$, where $\phi$ is the potential and $\vv$ is the peculiar 
velocity. We show that only two new free parameters are needed to model the 
power spectrum and bispectrum up to 4th order in the initial density 
perturbations, once symmetry considerations and equivalences between possible 
terms are accounted for. One of the new parameters is a bias multiplying 
$s_{ij}s_{ji}$, where $s_{ij}=\left[\nabla_i\nabla_j\nabla^{-2}-\frac{1}{3}
\delta^K_{ij}\right]\delta$. The other multiplies $s_{ij}t_{ji}$, where 
$t_{ij}=\left[\nabla_i\nabla_j\nabla^{-2}-\frac{1}{3}\delta^K_{ij}\right]
\left(\theta-\delta\right)$, with $\theta=-\left(a~H~d\ln D/d\ln a\right)^{-1}
\vnabla\cdot\vv$. (There are other, observationally equivalent, ways to write 
the two terms, e.g., using $\theta-\delta$ instead of $s_{ij}s_{ji}$.) We show 
how short-range (non-gravitational) non-locality can be included through a 
controlled series of higher derivative terms, starting with 
$R^2\nabla^2\delta$, where $R$ is the scale of non-locality (this term will be 
a small correction as long as $k^2 R^2$ is small, where $k$ is the observed 
wavenumber). We suggest that there will be much more information in future huge 
redshift surveys in the range of scales where beyond-linear perturbation theory
is both necessary and sufficient than in the fully linear regime. 
 
\end{abstract}


\maketitle

\section{Introduction}

\begin{figure}
\resizebox{\textwidth}{!}{\includegraphics{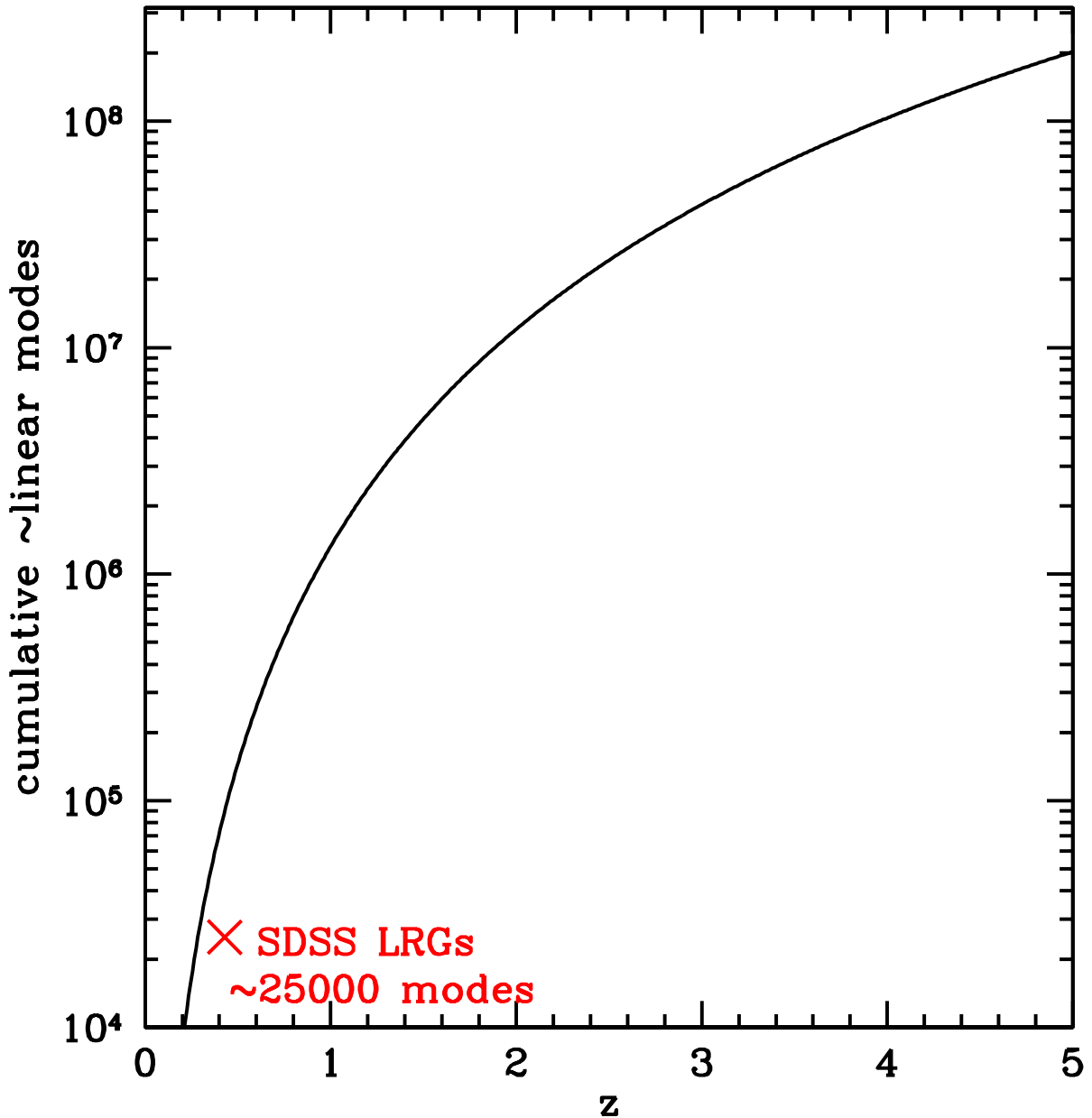}}
\caption{
Cumulative number of modes with $k<0.1/\left[D\left(z\right)/D\left(0\right)
\right]\ihmpc$ up to a given redshift.
The largest reasonably well-sampled LSS survey, the SDSS LRGs, probe only an
tiny fraction of the available modes.
}
\label{figmodes}
\end{figure}

While measurements of galaxy clustering have been around for a long time
\cite{1977ApJ...217..385G}, to 
the point where the casual observer might think they must 
surely be almost finished, or at least well-underway, in fact we have barely 
scratched the
surface of the possibilities for measuring large-scale structure (hereafter,
LSS, defined in this paper to mean surveys of any tracer of 
the large-scale mass density field -- we will often call the tracer 
``galaxies'', but it could just as
well be quasars \cite{2008JCAP...08..031S,2008arXiv0802.2105P}, the \lyaf\
\cite{2005ApJ...635..761M,2006ApJS..163...80M,2006MNRAS.365..231V},
galaxy cluster/Sunyaev-Zel'dovich effect measurements
\citep{2008RPPh...71f6902A},
21cm surveys \citep{2005MNRAS.364..743N,2008PhRvL.100i1303C}, etc.).
Measuring LSS should really be regarded as an exciting future   
probe of cosmology, with growth potential not {\it a priori} less than probes 
with less past success. 
The reason is simply that we have so far probed only a tiny fraction of the 
observable
volume of the Universe.  For example, the largest galaxy redshift survey with 
density approaching what is needed to fully sample the
near-linear regime of clustering, the Sloan Digital Sky Survey (SDSS) Luminous
Red Galaxy (LRG) survey \cite{2006PhRvD..74l3507T}, 
probes $\lesssim 2$ cubic Gpc/h, or 
$\sim 0.3$\% of the comoving volume at $z<5$.
Figure \ref{figmodes} shows that the fraction of linear regime modes, i.e., 
easily usable information, probed by the LRGs is even smaller --
barely 0.01\% of the modes at $z<5$ --
because the non-linear scale is smaller at higher $z$.
(For this figure, we
have used $k_{\rm NL}=0.1/\left[D\left(z\right)/D\left(0\right)
\right]\ihmpc$ for the non-linear scale, where $D$ is the linear growth factor.
The normalization 
$0.1 \ihmpc$ is somewhat arbitrary, depending on one's definition of 
the non-linear scale, but changing it only changes the 
overall normalization of the figure. The redshift dependence is motivated by 
\cite{2006PhRvD..73f3520C,2007ApJ...665...14S}.) 

The high precision of LSS statistics measured using future surveys 
probing appreciable fractions of the observable Universe 
\cite{2008arXiv0807.3140T,2008ExA...tmp...37C,2007AAS...21113229S,
2008arXiv0809.3002W,
2008PhRvL.100i1303C,2008arXiv0812.0419V,
2005MNRAS.360...27A,2008ASPC..399..115H,
2005astro.ph..7457G,2008A&G....49e..19B} will 
require an unprecedented level of accuracy in our
theoretical/phenomenological calculations of predictions for the statistics, 
if we are to fully exploit the potential of these surveys for measuring 
fundamental physics/cosmology. On very large scales we can use linear theory, 
but the scale below which linear theory cannot be trusted at the level of the 
error bars will become larger and larger (corresponding to a smaller and 
smaller maximum reliable wavenumber $k$) as the error bars shrink.
The number of Fourier modes in a
three-dimensional survey goes like the cube of the maximum usable $k$, i.e., 
in terms of raw information, extending the usable range of $k$ by a factor of
2 is equivalent to extending the volume of the survey by a factor of 8
(for a Gaussian field).  As we will see (Fig. \ref{figbasicgg}), the range of
scales where corrections to linear theory are small (perturbative), but still
statistically significant, can easily be a factor of $\sim 4$ for future large
surveys.  The point is simply that  
we have enormous leverage to extend the value of surveys through modeling 
improvements that extend the 
usable range of $k$.  For example, if a survey costs 50 million dollars, 
extending the effectively usable $k$ range by a mere factor of 1.3 (say, 
from $0.1\ihmpc$ to $0.13\ihmpc$) would be 
worth roughly 1000 person-years (at \$60000 per year).  
Phenomenological theory associated with LSS surveys should be viewed not as
a typical academic exercise, pursued by a few individuals or small groups 
because they think it is ``interesting'', 
but instead as an industrial, infrastructure building endeavor, critical to 
surveys in much the same way as, say, the road up to the telescope.

Better modeling is needed even for present, moderate precision surveys. 
For example, \cite{2008MNRAS.385..830S} shows clearly where the linear
bias model \cite{1984ApJ...284L...9K} 
that we have been relying on for cosmological parameter estimation
for decades is breaking down, by comparing 
results from SDSS and 2dF galaxies (see also \cite{2007ApJ...657..645P,
2008MNRAS.385.1635S}).
The power spectra of two different types of galaxies are
not related by a simple overall normalization factor (bias) -- their ratio
depends on
scale, even on quite large scales where it was once hoped that linear theory 
would be good enough.
This was not completely unanticipated, however,
\cite{2008MNRAS.385..830S} also shows that the ad hoc fitting formula of
\cite{2005MNRAS.362..505C}, that
has been used recently to try to account for quasi-linear galaxy clustering,
does not work well, and these problems lead to disagreement between
cosmological parameters inferred from different galaxy surveys
(see also \cite{2008JCAP...07..017H}).
Clearly, we have a lot of theoretical work to do if we want to fully exploit
future, much more precise, LSS data.

For measurements of the baryonic acoustic oscillation (BAO) feature
\cite{2007ApJ...665...14S,2007ApJ...664..660E,2007PhRvD..76f3009M,
2005ApJ...633..560E,2003ApJ...598..720S,1998ApJ...494L...1E,
1998ApJ...496..605E,2008arXiv0805.4238S,2008ApJ...686...13S,
2007MNRAS.377..185P}, ad hoc 
fitting formulas very carefully calibrated by simulations may be sufficient, 
but measuring other physics that
produces less distinctive signatures in the power spectrum, e.g.,
redshift-space distortions aimed at constraining dark energy 
\cite{2008arXiv0810.0323M,2008arXiv0810.1518W,2008arXiv0808.0003P,
2008JCAP...05..021W}, or 
measurements of the shape of the power spectrum aimed at constraining  
modified gravity \cite{2008arXiv0811.1272B,2008PhRvD..78d3514A}, 
neutrino masses 
\cite{2008PhRvD..78f5009P,2008PhRvD..77f3005K,2008JCAP...08..020B,
2008PhRvD..77h3507G,2008PhRvD..78c3010F,2007PhRvD..75e3001F,
2008PhRvL.100s1301S,2006PhRvD..73h3520T,2006PhR...429..307L,
2006PhRvD..73l3501S}, inflation \cite{2006PhRvD..73h3520T},
etc. \cite{2008arXiv0812.4016D,2008PhLB..663..160M,2006PhRvD..74d3505T}, 
will require well-motivated, rigorous descriptions of the relation
between galaxy and mass density, i.e., bias models. 
In other words, better LSS theory will substantially enhance the constraining 
power of BAO-oriented surveys, by allowing the use of non-BAO information
\cite{2008arXiv0805.2632J}.

Bias modeling can be roughly divided into two approaches
(excluding attempts to simulate galaxies from something 
resembling first principles \cite{2001ApJ...558..520Y,1999ApJ...522..590B},
which can be useful as a guide/spot-check for other methods,
but are unlikely to be accurate and
efficient enough to use for interpretation of precision statistics any time
soon):
The first approach might be called a bottom-up approach, 
where one starts with a model for how individual galaxies sit in 
the local small-scale mass
density field (most recently almost always based on galaxies sitting in dark 
matter halos, but
earlier on peaks or other features), and then computes large-scale clustering
by including the large-scale correlation of the relevant small-scale 
density feature. The other approach might be called top-down, or perturbative,
where one starts from the fact that large-scale fluctuations are 
small and expands a completely unknown relation between galaxies
and mass, with generally
infinite freedom (except typically for the assumption of locality, relative 
to the scale 
of observations) into a Taylor series in the density perturbations, where the 
coefficients of the first few
terms in the series become the free parameters of the model 
(the main point of the renormalized bias scheme 
of \cite{2006PhRvD..74j3512M} was to demonstrate how
this separation of scales can be done in an organized way ---
see \cite{2002PhR...367....1B} for a general review of LSS perturbation 
theory). 

This paper takes the perturbative approach, but 
most recent work has been based in some way on dark matter halos (e.g.,
\cite{2008arXiv0808.2988Y,2008arXiv0812.4288W,2007ApJ...659....1Z,
2006ApJ...652...26Y,2005ApJ...631...41T,2005MNRAS.362..337N,
2000MNRAS.318..203S,2001MNRAS.325.1359S,2004MNRAS.355..129S,
2000ApJ...542..559T,1997MNRAS.284..189M,1996ApJS..103...63B,
1996ApJS..103....1B,1996ApJS..103...41B}).  
A strong foundation for halo  models is the expectation that, with enough work, 
it should be possible to make accurate numerical simulations of 
the large-scale clustering of halos within a given cosmological model 
\cite{2008arXiv0804.0004R}
(it is much more difficult to fully quantify this
clustering to the point where one does not need to make halo models based on
the halos in full simulations, but that is only necessary for convenience). 
Unfortunately, we can see these halos only through the coarse 
probe of gravitational lensing \cite{2005PhRvD..71d3511S}, and it is 
not straightforward to determine the relation between halos and the more easily 
observable galaxies. 
The halo models therefore specify a ``halo occupation distribution'' (HOD) for
the galaxies, i.e., a recipe for populating halos with galaxies.
The hope of these models is that they can determine the HOD using 
information deeper into the non-linear 
regime than possible using the more general, less predictive,
perturbative approach that we will discuss, but this is a difficult game.
To be reliable, 
models that populate halos within a full numerical simulation must 
include enough freedom in the method for populating halos to cover all 
realistic possibilities.
Models that further rely on analytic calculations for the clustering of halos 
introduce another level of complexity and possibility of error 
\cite{2008PhRvD..78b3523S,2007PhRvD..75f3512S}.

To appreciate the small-scale complexity that we will bundle into a few
perturbative bias parameters, it is useful to review the recent work toward 
understanding the details of halo models.
The standard HOD assumption is that the number of galaxies in a halo is 
some relatively simple function of the mass of the halo.
Even these relatively simple HODs have $\sim 10$ free parameters
\cite{2007ApJ...659....1Z}. 
There is observational evidence that this form of HOD works 
qualitatively very well \cite{2008ApJ...686...53T}; however, the assumptions
involved clearly can not be perfect. 
\cite{2005MNRAS.363L..66G} showed that the clustering 
of halos of a fixed mass depends significantly on the time when the halo formed
(see also \cite{2006ApJ...652...71W,2007MNRAS.378..777R,
2006MNRAS.367.1039H}). 
This phenomenon is often called assembly bias.
When combined with the possibility that the galaxy population
within halos of a given mass can depend on the halo formation time,
this means that it is necessary for the HOD to depend on more parameters than 
just mass.  
\cite{2007MNRAS.374.1303C} demonstrated this explicitly using semi-analytic
models for galaxy formation (see also \cite{2008ApJ...686...41Z}), and found 
that accounting for formation time 
or halo concentration in addition to mass explains only a fraction of the 
effect.
\cite{2008MNRAS.389.1419L} found that the magnitude and mass-dependence of the
assembly bias depends on the definition of halo formation time (different 
definitions capture different aspects of the history of the halo).
\cite{2008MNRAS.387..921A} extends these results to higher order statistics.
\cite{2007ApJ...656..139W} showed that the clustering of massive halos depends
on concentration in addition to mass, and also recent history of mergers.
The simulations of \cite{2007MNRAS.375..633W,2008arXiv0803.4211H} suggest that
the relation between formation time and clustering for small halos
is due to the effect of tides in high density regions suppressing later 
growth of small halos.
The simulations and analytic calculations of \cite{2008ApJ...687...12D} 
suggest that 
at low masses assembly bias is 
again related to high density regions suppressing late-time accretion, and
at high masses the effect is related to the curvature around
the initial peak that grows into the halo.
The simulations of \cite{2008arXiv0811.3214D} show that the clustering of halos
at high redshift also depends significantly on their angular momentum, at fixed
mass.
Finally, simulations even show a population of halos that were once subhalos 
within
a larger halo, but were ejected by interactions \cite{2008arXiv0811.3558W}.
Not surprisingly, the ejected halos do not cluster in the same way as other 
halos of the 
same mass. Generally, the idea that the mass density field breaks up neatly
into halos, containing galaxies, which retain little information about their 
formation process, is
a great qualitative way to picture the formation of structure, but we should
not forget that it is a picture, not a calculation.
Another assumption of typical halo models is that
the distribution of satellite galaxies within dark matter halos 
follows the mass density profile, but this has been only roughly 
justified \cite{2006MNRAS.366.1529M,2005MNRAS.356.1233V,
2005ApJ...618..557N,2003ApJ...593....1B}.
Explanations of why these issues are not fundamental problems for the HOD 
approach make the argument that the effects are not large enough to matter
now, but not that they will not in the future \cite{2007ApJ...659....1Z}. 

In the face of {\it any} uncertainty about whether the
small-scale halo model is 
sufficient, a precision measurement of fundamental physics/cosmology that is 
consistent with prior expectations may be believed,
but a truly new, unexpected result will not be. This is only a very meager 
form of progress.
The same kind of thing can be said about the perturbative
approach -- as long as there is any question of whether the bias description
is complete, the results will not be believed in any important situation.
We believe that it is reasonable to hope that the 
perturbative bias approach can be made relatively airtight, as long as
one does not try to push it beyond its range of 
validity. This paper is an attempt to make progress in that direction.

General understanding of large-scale clustering, independent of specific 
small-scale models for the dark matter tracer, has been developing gradually.
\cite{1993MNRAS.262.1065C} showed that if the galaxy density is a general 
function of the local mass density, and the mass density field is assumed to 
be Gaussian, the asymptotically large-scale galaxy correlation function will be 
proportional to the mass correlation function (except for special cases of
the local function). \cite{1993MNRAS.262.1065C} also showed that, under the
same conditions, the galaxy
power spectrum may go to a constant as $k\rightarrow 0$ (even if no white noise
is introduced by hand).
\cite{1993ApJ...413..447F} introduced the perturbative bias model in the form
that we will follow, where the
galaxy density perturbation $\delta_g$ is first written as a completely general
function,
$f(\delta)$, of the mass density perturbation $\delta$, and then the function 
is Taylor
expanded, with the unknown coefficients in the series becoming the bias
parameters, $b_i$, i.e., 
\begin{equation}
\delta_g(\vx)=f(\delta(\vx))=\sum_{i=0}^\infty \frac{b_i}{i!}\delta^i(\vx)~,
\end{equation}
with the mass density given by gravitational perturbation theory.
Note that the observation that the first order term in this series describes 
simple
scale-independent linear bias does not guarantee that higher order terms 
cannot cause large-scale deviations from this form.
\cite{1998ApJ...504..607S} showed, starting with the same Taylor series form of
bias, that if the mass clustering is 
hierarchical, then $\xi_g(r)\propto \xi(r)+\mathcal{O}(\xi^2)$, even if the 
local
bias relation is applied on scales where the fluctuations are not small.
The large-scale bias factor found by \cite{1998ApJ...504..607S} was an 
infinite sum of terms proportional to powers of the mass density variance,
a foreshadowing of the renormalized bias approach we follow in this paper
\cite{2006PhRvD..74j3512M}.
They went on to show that the linear bias relation holds even if the local 
mass density does not determine the galaxy density uniquely, but only 
determines a random distribution for the galaxy density (with the randomness
in that distribution independent from point to point).
Finally, \cite{1998ApJ...504..607S} showed that the galaxy power spectrum 
obeys the linear bias relation on scales similar to the correlation function,
except the small-separation part of 
the correlation function, which deviates from linear bias, will contribute 
an added constant to the power spectrum (see also 
\cite{1999ApJ...520...24D,2003ApJ...585L...1D,1999ApJ...511....1S}),
a foreshadowing of the noise renormalization that we will employ
\cite{2006PhRvD..74j3512M}. \cite{1998MNRAS.301..797H} found similarly that 
higher order corrections in straightforward gravitational perturbation theory 
starting from
the local Taylor series model for bias produce terms that on large scales
look like modifications of the linear theory bias or additional shot-noise.
Generally, it has been pretty well established that  
linear bias plus white noise is the correct model for very 
large scale galaxy clustering \cite{1999ApJ...525..543M,2000ApJ...528....1N,
1999ApJ...521L...5C},
barring the introduction of long-range non-gravitational effects which
essentially introduce deviations from this form by hand. 
\cite{2006PhRvD..74j3512M} put these results together into a neat 
computational
package, by employing renormalization ideas from quantum field theory
\cite{Peskin:1995ev} (some similar ideas were present in 
\cite{1999ApJ...522...46T}). The inconvenient results of 
\cite{1998ApJ...504..607S,1998MNRAS.301..797H}, that higher order calculations 
can affect clustering statistics on arbitrarily large scales, and that these
corrections are sensitive to the assumed small scale smoothing (cutoff),
are rendered observationally irrelevant by absorbing the inconvenient pieces 
into renormalizations of the existing bias parameters (including the noise 
level). 
This approach clears the way for pushing, in a systematic way, beyond
the very large-scale, purely linear, regime and into the information-rich 
smaller 
scales where higher order corrections are non-negligible, and understanding
the smoothing/cutoff issue becomes critical.
\cite{2008arXiv0805.2632J} showed that this approach describes clustering in 
simulations very well.

Remarkably, for all of the work on both the halo-based and perturbative 
approaches to bias, neither have generally been 
adopted, beyond the papers in which they are proposed, for use
in the main stream of LSS power spectrum measurement and  
cosmological parameter estimation \cite{2008arXiv0803.0547K,
2006PhRvD..74l3507T,
2007ApJ...657..645P,2006JCAP...10..014S}. 
In fact, even the proposers generally have not pushed 
their methods through to the point of making comprehensive parameter 
measurements (see \cite{2005ApJ...625..613A} for an exception).
The widespread use of the demonstrably inadequate (when extrapolated 
beyond 
its original purpose) fitting formula of \cite{2005MNRAS.362..505C}
should really be seen as an embarrassing failure of the LSS theory 
community. This paper will, unfortunately, continue this legacy of failure, 
but with the hope that it can soon be rectified.

In this paper, we will 
improve the Eulerian bias model by allowing for dependence on the local 
velocity
divergence and shear and the tidal tensor in addition to density. The reason to
expect such
dependence at some level is simple: two patches of space with the same final 
density did not necessarily follow the same path to reach that density, and
that difference in history may affect the galaxy density at the time of 
observation. 
In perturbation theory up to some finite order, however, the 
entire density history of 
a patch is reconstructible given a finite number of local quantities like the
the velocity divergence and tidal tensor. This raises the hope that a
completely unique, general, bias model can be constructed, covering all 
possibilities
for large-scale clustering with a finite set of bias parameters. (One can
always imagine unavoidable obstacles to this, e.g., long-range 
non-gravitational effects like inhomogeneous reionization affecting clustering
\cite{2006ApJ...640....1B,2007JCAP...10..007C},
however, to the extent that something like this is important 
on a given scale,
very high precision cosmology is probably simply impossible on that scale.) 
While the primary philosophy of this paper is that any possible form of 
large-scale clustering should be included in the model, unless it can be 
compellingly rejected, there is actually a lot of evidence that these new
forms of bias are needed, related to the
assembly bias phenomenon seen in simulations
\cite{2008arXiv0803.4211H} or observational correlations between galaxy 
properties and their environment \cite{2008arXiv0803.1759L}.

In a very interesting paper,
\cite{2008PhRvD..78j9901M,2008PhRvD..78h3519M} 
points out that a perturbative bias model assumed to 
be local in initial Lagrangian density produces results distinct from the model 
assumed to be local in final Eulerian density
(see also \cite{2000MNRAS.318L..39C,1998MNRAS.297..692C}). 
While \cite{2008PhRvD..78j9901M}
presents this as an advantage of Lagrangian PT, which is supposed to be a more
correct way to look at bias, we believe that it is better to
say that this represents a deficiency in the development of one or both 
approaches, not a conceptual problem with either. 
As a first approximation, it may be more accurate to assume that
bias is local in the initial Lagrangian density than the final Eulerian 
density, but neither assumption can be rigorously justified. 
Barring the unlikely proof that one approach is fundamentally 
superior to the other,
one criteria for believing future very high precision cosmology measurements
should be that
Lagrangian and Eulerian PT give equivalent answers in regimes where 
the calculations
converge, once all possible freedom is included in each version of the bias 
model. We prefer to work with the Eulerian model simply because it is 
expressed in terms of quantities that are generally more directly observable.  
This paper will implicitly address the differences between Lagrangian and
Eulerian PT raised by \cite{2008PhRvD..78j9901M,2008PhRvD..78h3519M}.

Note that, while we primarily discuss results in terms of the power spectrum,
nothing about the perturbative approach intrinsically requires one to go to
Fourier space. It is
simple to obtain the correlation function by Fourier transforming the power
spectrum, but it is also possible
to do all of the same calculations, from scratch, in configuration space.

The plan of the rest of the paper is as follows: 
In \S \ref{secbasicmodel} we discuss the
primary new extensions to the Eulerian bias model that we will work out fully
in this paper: including dependence on the local large-scale tidal tensor
and velocity divergence and shear. 
In \S \ref{secextensions} we briefly discuss some further 
extensions that are implied by the same line of thinking, related to 
redshift-space distortions, short-range non-locality, and non-Gaussianity of 
the primordial perturbations, although we will 
not fully develop them. Finally, in \S \ref{secconclusions} we will give some
conclusions and thoughts on directions for future work. 

\section{A more general Eulerian bias model \label{secbasicmodel}}

In this section we lay out a baseline extension to the model of galaxy bias
as dependent on local density only. In \S \ref{secvariables} we discuss the 
variables we will allow the galaxy density to depend on, 
and in \S \ref{secstatistics}
we compute statistics of galaxy clustering using these variables.

\subsection{Independent variables \label{secvariables}}

This subsection seeks to answer the question:
In general, in principle, in perturbation theory, what can the galaxy density 
depend on?

Everything we know about LSS at a given time in standard
perturbation theory (PT) is contained in the dynamical variables 
$\delta\left(\vx\right)=\rho_m\left(\vx\right)/\bar{\rho}_m-1$, where 
$\rho_m\left(\vx\right)$ is the mass density at position $\vx$ and 
$\bar{\rho}_m$ is the mean mass density, and 
$\theta\left(\vx\right)=\vnabla\cdot \vv\left(\vx\right)$, where $\vv$ is the 
peculiar velocity (see \cite{2002PhR...367....1B} for a review of LSS PT --
note that we will make the usual approximation that the Einstein-de Sitter PT 
results can be used for other models as long as the linear 
growth factor is replaced by the growth factor in the desired model).  
Because the velocity field is curl-free, it can be derived from 
$\theta$, i.e., $v_i = \nabla_i~ \nabla^{-2} \theta$ 
($\nabla^2\equiv \nabla_i\nabla_i$, and 
$\nabla^{-2}$ represents the usual $r^{-1}$ potential integral, or $-k^{-2}$ in 
Fourier space).
To allow for the non-locality (in the density field) introduced by 
gravity, we
will also consider dependence of the galaxy density on the local potential 
field, $\phi(\vx)$, which
can always be derived from $\delta$ using the Poisson equation. Allowing 
dependence
on $\vv(\vx)$ and $\phi(\vx)$, in spite of the fact that the system is 
entirely determined by $\delta(\vx)$ and $\theta(\vx)$, can be understood
as allowing for history dependence of the number of galaxies in a given patch 
of space, i.e., these
quantities tell us about the path the patch took to get to the density and 
velocity divergence that it has. 

A homogeneous change in $\phi$ should not be observable, which suggests that
the galaxy density should only depend on $\nabla_i \phi$.
Furthermore, a homogeneous gravitational force shouldn't be observable either,
suggesting that we should use $\nabla_i \nabla_j \phi$.  Therefore we define:
\begin{equation}
s_{i j}\left(\vx\right)\equiv
\nabla_i \nabla_j \phi\left(\vx\right) - \frac{1}{3} \delta^K_{ij}~ 
\delta\left(\vx\right)=
\left[\nabla_i \nabla_j\nabla^{-2} - \frac{1}{3} \delta^K_{ij}\right]
\delta\left(\vx\right)\equiv \gamma_{ij}\delta\left(\vx\right)~,
\end{equation}
where we have removed the trace of $\nabla_i \nabla_j \phi$ because it is
redundant with $\delta$ (note that we are absorbing all of the spatially 
constant factors in the Poisson equation into the definition of $\phi$, 
i.e., $\nabla^2\phi \equiv \delta$ -- we will make a similar re-definition of 
$v_i$ to make $\theta \equiv \delta$ in linear theory).
For compactness, we have defined the operator
\begin{equation}
\gamma_{ij}\equiv \nabla_i \nabla_j~ \nabla^{-2}-
\frac{1}{3} \delta^K_{ij} ~.
\end{equation}  
Similarly, a homogeneous velocity field should not be observable, suggesting 
that galaxy density depends on velocity through 
$\nabla_i v_j=\nabla_i \nabla_j~ \nabla^{-2} \theta$.  Because $\theta=\delta$
at linear order, $\nabla_i v_j$ is redundant with $\nabla_i \nabla_j \phi$ at
linear order, so it simplifies things in perturbation theory to use their 
difference for our independent variables, i.e., to define
\begin{equation}
\ttheta\left(\vx\right)\equiv \theta\left(\vx\right) - \delta\left(\vx\right) ~,
\end{equation}
and 
\begin{equation}
t_{i j}\left(\vx\right)\equiv 
\nabla_i v_j\left(\vx\right)-\frac{1}{3} \delta^K_{ij}\theta\left(\vx\right) 
-s_{ij}\left(\vx\right)=
\left(\nabla_i \nabla_j~ \nabla^{-2}-\frac{1}{3} \delta^K_{ij}\right) 
\left[\theta\left(\vx\right)-
\delta\left(\vx\right)\right]
=
\gamma_{ij}\ttheta\left(\vx\right)~.
\end{equation}
The difference variables $\ttheta$ and $t_{ij}$ are non-zero only at 2nd order.

Now, the galaxy density will depend on $\delta$, $s_{ij}$, $\ttheta$, and 
$t_{ij}$, but it
can't depend directly on anything but a scalar quantity.
This is because, assuming homogeneity and isotropy, we can only have 
constant, scalar, bias parameters.  For example, the general Taylor series for 
a function that 
depends on a small tensor $\sigma_{ij}$ is
\begin{equation}
f\left(\sigma_{ij}\right)=f\left(0\right)+
\frac{df}{d \sigma_{ij}}\left(0\right)
\sigma_{ij}+...\equiv p_0+p_{ij} \sigma_{ij}+...~.
\label{eqtensorTaylor}
\end{equation}
In general, each element of $p_{ij}$ could be 
independent, but this is inconsistent with isotropy.  The only 
consistent possibility is $p_{ij}=p_1 \delta_{ij}^K$.  In this case, only 
$\sigma_{ii}$ enters the Taylor series.  Similar arguments apply to higher 
order terms.

By construction, $s_{ii}=0$
and $t_{ii}=0$.  We can construct products, up to 3rd order in the initial
perturbations, $s^2\equiv s_{ij} s_{ji}$, 
$st\equiv s_{ij} t_{ji}$, and 
$s^3 \equiv s_{ij} s_{jk} s_{ki}$ ($t_{ij} t_{ji}$ is 4th order). 
It turns out that, at 2nd order in PT, 
$\ttheta_2=\frac{2}{7}s_1^2-\frac{4}{21}\delta_1^2$.
This suggests that, in place of $\ttheta$, we use a variable 
constructed to be zero at both 1st and 2nd order in standard PT,
\begin{equation}
\tttheta \left(\vx\right)\equiv \ttheta\left(\vx\right)-
\frac{2}{7} s^2\left(\vx\right) +\frac{4}{21}\delta^2\left(\vx\right)~.
\end{equation}
This definition makes $\tttheta$ non-zero only at 3rd order.
Note that we can not redefine $t_{ij}$ in terms of $\tttheta$ because this would
require terms like $\gamma_{ij}\delta^2$.
To summarize, our galaxy density will (naively) be a Taylor series involving 
the following eight quantities:
\begin{eqnarray}
{\rm 1st~ order}&:&\delta \\ \nonumber
{\rm 2nd~ order}&:& \delta^2,~ s^2 \\ \nonumber
{\rm 3rd~ order}&:& \delta^3,~ \delta s^2,~\tttheta,~st,~s^3 \\ \nonumber
\end{eqnarray}
This shows why standard linear theory bias, $\delta_g = b~\delta$, is
sufficient in the truly linear regime:  all other independent scalar 
quantities we can form are higher order.

Finally, our model, which now starts with
$\rho_g=f(\delta,~\nabla_i\nabla_j\phi,~\nabla_i v_j)$, will be extended to 
include general dependence on a mean-zero Gaussian white noise variable 
$\epsilon$, 
i.e., $\rho_g=f(\delta,~\nabla_i\nabla_j\phi,~\nabla_i v_j,~\epsilon)$, 
to allow for
stochasticity and shot-noise in the galaxy density-mass density relation.
This approach 
is new relative to past work where a noise variable was simply
tacked onto the end of the Taylor series. We will Taylor expand around 
$\epsilon=0$, just like the other variables, treating epsilon as similar
in size to $\delta$, and including all higher order terms. 
This may appear strange, and actually will not affect power spectrum 
calculations at all, but we will see when we compute the bispectrum that this
is a compact way to include the fact that Poisson sampling of the density field
actually affects the bispectrum, in contrast to Gaussian noise
\cite{1980lssu.book.....P,2008PhRvD..78b3523S}.

A Taylor series in these quantities, up to 3rd order in the initial 
perturbations, is
\begin{eqnarray}
\rho_g&=& 
p_0 + 
p_\delta~\delta + 
\frac{1}{2}~ p_{\delta^2}~ \delta^2 +
\frac{1}{2} p_{s^2}~ s^2 + 
\frac{1}{3!}~ p_{\delta^3}~ \delta^3 +
\frac{1}{2}p_{\delta s^2}~ \delta~ s^2 +
p_{\tttheta}~ \tttheta+ 
p_{st}~ st+
\frac{1}{3!}~ p_{s^3}~ s^3 
\\ \nonumber
& &+ p_\epsilon~\epsilon + 
p_{\delta\epsilon}~\delta\epsilon + 
\frac{1}{2}~ p_{\delta^2\epsilon}~ \delta^2\epsilon +
\frac{1}{2} p_{s^2\epsilon}~ s^2\epsilon + 
\frac{1}{2} p_{\epsilon^2}~\epsilon^2 + 
\frac{1}{2} p_{\delta\epsilon^2}~\delta\epsilon^2 + 
\frac{1}{3!} p_{\epsilon^3}~\epsilon^3 + 
...
\end{eqnarray} 
(note that the factors of $1/2$ and $1/3!$ serve no real purpose, because the
$p$'s are essentially arbitrary and could be redefined to include these 
factors).

One might ask at this point:  Why not add more derivatives, e.g., terms like
$\nabla^2\delta$ or products of $\gamma_{ij}\gamma_{kl}~\delta$? 
Also, why not make the dependence non-local, i.e.,
\begin{equation}
\rho_g\left(\vx\right)=f\left[\delta\left(\vx^\prime\right)\right]~,
\end{equation}
where $\vx^\prime$ can be any position, not just the position $\vx$ where we
are measuring the density.
It turns out that these things are related, as we will discuss further in 
\S \ref{secnonlocal}.  As long as the non-locality is short range, it can
be easily represented by a controlled series of higher derivative terms like 
$\nabla^2\delta$.
Terms like $\gamma_{ij}\gamma_{kl}\delta$, which we will not consider, 
introduce new long-range
$\nabla^{-2}$ operators, beyond the one already present in the construction of 
the gravitational potential. 

One might also wonder about the eigenvalues of $s_{ij}$, $\lambda_i$
\cite{2008PhRvD..78b3527D}: Are they 
not additional scalar 
quantities that are linear in the perturbation amplitude, and thus loopholes
in the argument that linear order bias can only depend on $\delta$? In three 
dimensions,
they are hard to write down explicitly, but the two dimensional version is 
informative:
$\lambda_{\pm}=\pm \sqrt{\frac{1}{4}\left(s_{11}-s_{22}\right)^2+s_{12}^2}$.
We see that these quantities are in some sense the same order as $\delta$, but
they are not well behaved analytic functions of $s_{ij}$. 
This is illustrated by considering a similar, but simpler to understand,
possible term,  
$\left|\delta\right|=\sqrt{\delta^2}$. At $\delta=0$, $\left|\delta\right|$ is 
not differentiable, and it becomes especially obvious how unphysical
this must be when we observe that local physics has no particular reason to 
see the mean density of the Universe as a special value. Similarly, it seems
unlikely that it is physically correct for the dependence of galaxy density on
$s_{11}-s_{22}$ (for $s_{12}=0$) to make a sharp change of direction at 
$s_{11}-s_{22}=0$ (which is just the transition from a tensor extended in the
1 direction to the 2 direction), as it would if we included terms linear in 
the eigenvalues. It is undoubtedly possible for the galaxy density to depend
on these eigenvalues -- the argument here is simply that this dependence should
be higher than linear order. Our 
parameterization actually already includes this dependence very directly: 
$s^2=s_{ij} s_{ji}=\lambda_i \lambda_i$, i.e., $s^2$ is the sum of squares of
the eigenvalues.

The bottom line is: We stick to the terms that are obtained in a Taylor series
in $\delta$, $\partial_i v_j$, and $\partial_i \partial_j \phi$, with only 
short range (relative to the scale of observations) non-locality in the 
dependence of galaxy density on these quantities.
We leave for the future the question of how completely general this approach 
is.

\subsection{Statistics \label{secstatistics}}
The mean galaxy density is, to 3rd order in the initial perturbations,
\begin{equation}
\bar{\rho}_g\equiv \left<\rho_g\right>=p_0+\frac{1}{2}p_{\delta^2}~ \sigma^2
+\frac{1}{3}p_{s^2}~\sigma^2+\frac{1}{2} p_{\epsilon^2}~\sigma^2_\epsilon~,
\end{equation}
where $\sigma^2=\left<\delta^2\right>$, 
$\left<s^2\right>=\frac{2}{3}\sigma^2$, and 
$\sigma^2_\epsilon=\left<\epsilon^2\right>$.
Redefining all the coefficients after division by $\bar{\rho}_g$ gives
\begin{eqnarray}
\delta_g&\equiv& \rho_g/\bar{\rho}_g-1 \\ \nonumber
&=& c_\delta~\delta + 
\frac{1}{2}~ c_{\delta^2}~\left( \delta^2-\sigma^2\right) +
\frac{1}{2} c_{s^2}~ \left(s^2-\frac{2}{3}\sigma^2\right) + 
\frac{1}{3!}~ c_{\delta^3}~ \delta^3 +
\frac{1}{2}c_{\delta s^2}~ \delta~ s^2 +
c_{\tttheta}~ \tttheta+ 
c_{st}~ st+
\frac{1}{3!}~ c_{s^3}~ s^3
\\ \nonumber
& &+ c_\epsilon~\epsilon + 
c_{\delta\epsilon}~\delta\epsilon + 
\frac{1}{2}~ c_{\delta^2\epsilon}~ \delta^2\epsilon +
\frac{1}{2} c_{s^2\epsilon}~ s^2\epsilon + 
\frac{1}{2} c_{\epsilon^2}~\left(\epsilon^2-\sigma_\epsilon^2\right) + 
\frac{1}{2} c_{\delta\epsilon^2}~\delta\epsilon^2 + 
\frac{1}{3!} c_{\epsilon^3}~\epsilon^3 + 
...
\end{eqnarray}

\subsubsection{Galaxy-mass cross-spectrum}

For simplicity, we start by calculating the mass density-galaxy density
cross-spectrum, i.e., $\left<\delta_m\left(\vk\right)\delta_g
\left(\vk^\prime\right)\right>=
\left(2\pi\right)^3 \delta^D\left(\vk+\vk^\prime\right) P_{mg}(k)$, which is
\begin{eqnarray}
\label{eqmassgal}
P_{mg}(k)&=&
c_\delta~P_{\rm NL}(k) \\ \nonumber 
&+& 
c_{\delta^2}~\int \frac{d^3\vq}{\left(2 \pi\right)^3}P\left(q\right) 
P\left(\left|\vk-\vq\right|\right)\FtwoS\left(\vq,\vk-\vq\right)+
 \frac{34}{21}~ c_{\delta^2}~\sigma^2~ P\left(k\right)
\\ \nonumber &+& 
c_{s^2}~\int \frac{d^3\vq}{\left(2 \pi\right)^3}P\left(q\right)
P\left(\left|\vk-\vq\right|\right)
\FtwoS\left(\vq,\vk-\vq\right)
\ssqfac\left(\vq,\vk-\vq\right)   \\ \nonumber
&+&2~ c_{s^2}~P\left(k\right)~ 
\int \frac{d^3\vq}{\left(2 \pi\right)^3}P\left(q\right)
\FtwoS\left(-\vq,\vk\right)
\ssqfac\left(\vq,\vk-\vq\right)
\\ \nonumber &+&
\frac{1}{2}~ c_{\delta^3}~ \sigma^2~ P\left(k\right) +
\frac{1}{3}~ c_{\delta s^2}~ \sigma^2~ P\left(k\right)
\\ \nonumber&+&
2~ c_{\tttheta}~P(k)~ \int \frac{d^3\vq}{\left(2 \pi\right)^3}P\left(q\right)
\left[\frac{3}{2}D_S^{\left(3\right)}\left(\vq,-\vq,-\vk\right)-2~
\FtwoS\left(-\vq,\vk\right)
D_S^{\left(2\right)}\left(\vq,\vk-\vq\right)
\right]  
\\ \nonumber &+&
2~ c_{st}~ ~P(k)~ \int \frac{d^3\vq}{\left(2 \pi\right)^3}
P\left(q\right) D_S^{\left(2\right)}\left(-\vq,\vk\right)
\ssqfac\left(\vq,\vk-\vq\right) 
\\ \nonumber &+&
\frac{1}{2}~c_{\delta\epsilon^2}~\sigma^2_\epsilon~ P(k) 
~.
\end{eqnarray} 
See the Appendix for definitions of $F_S$, $S$, and $D_S$.
$P_{\rm NL}(k)$ is the non-linear mass power spectrum.
$P(k)$ with no subscript always refers to the linear theory mass power.
Note that the $s^3$ term works out to exactly zero, so the parameter 
$c_{s^3}$ has been rendered irrelevant.

As we found in \cite{2006PhRvD..74j3512M}, some terms like 
$\frac{1}{2} c_{\delta^3} \sigma^2 P\left(k\right)$ appear which are best 
treated as renormalizations of the linear theory bias, i.e., by a redefinition
like $c_\delta^\prime=c_\delta+\frac{1}{2} c_{\delta^3} \sigma^2$.
As discussed in \cite{2006PhRvD..74j3512M}, the un-smoothed 
density variance $\sigma^2=\left<\delta^2\right>$ may not be literally 
infinite, depending on the power spectrum, but it will be large, and sensitive
to the deeply non-linear regime where all of our calculations are meaningless.
It is best to think of the original $c_\delta$ as an un-observable ``bare''
parameter, with the observable linear bias factor being largely un-related to 
it as the sum of many higher order terms which are generally much larger. 
This idea that the values of the parameters of large-scale galaxy clustering 
are generated
by small-scale, higher order effects is physically reasonable, or even
expected --- after all, if there were truly only small, linearizable,
perturbations in the Universe, there would be no galaxies.

\begin{figure}
\resizebox{\textwidth}{!}{\includegraphics{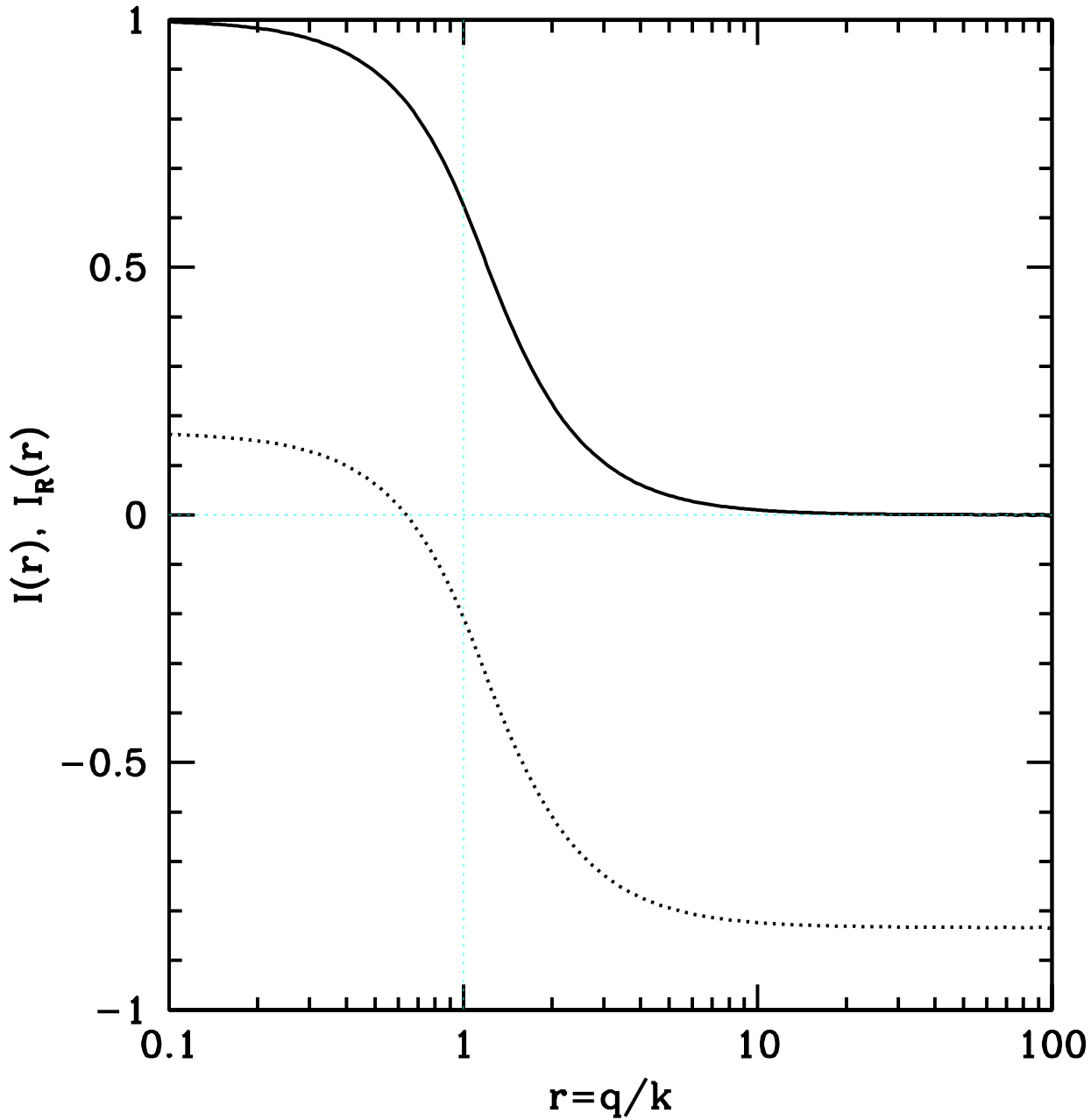}}
\caption{
Weighting kernel over which $\Delta^2\left(q=r~k\right)$ is integrated to 
obtain the contribution of several terms to 
$P_{mg}\left(k\right)$.  The dotted line shows $I\left(r\right)$, 
defined by Eq. (\ref{eqIdef}), which is sensitive to high-$k$ power.
The solid line shows the kernel after renormalization of the linear bias, 
$I_R\left(r\right)
=I\left(r\right)+5/6$, which now acts as a filter to produce the variance of
the density field smoothed on scale $~k$. 
}
\label{figstexamp}
\end{figure}

The term associated with $st$ has an interesting new feature.
In the $k\rightarrow 0$ limit, we find
\begin{equation}
2~ c_{st}~P(k)~ \int \frac{d^3\vq}{\left(2 \pi\right)^3}P\left(q\right)
D_S^{\left(2\right)}\left(-\vq,\vk\right)
\ssqfac\left(\vq,\vk-\vq\right)
~\stackrel{k \rightarrow 0}{\longrightarrow}~ 
-\frac{16}{63}~ c_{st}~\sigma^2~ P(k) ~.
\end{equation}
Like the 
$\delta^3$ term, for example, this looks like a renormalization of the linear 
bias; however,
unlike the $\delta^3$  term, here there is non-trivial $k$ dependence as one 
goes to non-zero $k$.
This case provides an opportunity to demonstrate how the renormalization 
works more clearly.  
Defining $r=q/k$, $\mu=\vk\cdot \vq/k~q$, and
\begin{equation}
I\left(r\right)=\frac{105}{32}
\int_{-1}^1 d\mu~ D^{\left(2\right)}\left(-\vq,\vk\right)
~\ssqfac\left(\vq,\vk-\vq\right)
\label{eqIdef}
\end{equation} 
we have
\begin{equation}
2~ \int \frac{d^3\vq}{\left(2 \pi\right)^3}P\left(q\right)~
D^{\left(2\right)}\left(-\vq,\vk\right)~
\ssqfac\left(\vq,\vk-\vq\right)=
\frac{32}{105}
\int d\ln r~ \Delta^2\left(k r\right)~ I\left(r\right)
\end{equation}
where $\Delta^2\left(q\right)\equiv q^3 P\left(q\right)/2 \pi^2$.
$I\left(r\right)$ gives the weight function over which one must integrate 
$\Delta^2\left(q\right)$ to obtain the bias term.
Figure \ref{figstexamp} shows a plot of $I\left(r\right)$.
We see that $I\left(r\right)$ is constant as $r\rightarrow \infty$.  This
leads to the constant result as $k\rightarrow 0$, and is clearly 
undesirable as it represents sensitivity to arbitrarily small, highly 
non-linear scales.  
The solution is to subtract the $k\rightarrow 0$ result, i.e., 
$-\frac{16}{63}~ c_{st}~\sigma^2~ P\left(k\right)$, from this term,
and add it to the linear theory bias.
The remainder is
\begin{equation}
2~ c_{st}~P(k)~ \int \frac{d^3\vq}{\left(2 \pi\right)^3}P\left(q\right)
\left[
D^{\left(2\right)}\left(-\vq,\vk\right)
\ssqfac\left(\vq,\vk-\vq\right)+\frac{8}{63}\right]=
c_{st}~P(k)~\frac{32}{105}
\int d\ln r~ \Delta^2\left(k r\right)~ I_R\left(r\right)
\end{equation}
where
\begin{equation}
I_R\left(r\right)=I\left(r\right)+5/6~,
\end{equation}
$I_R$ now 
looks like a smoothing kernel, with no sensitivity to power for 
$r>>1$, i.e., $q>>k$
(the factor $105/32$ was chosen
to make $I_R\left(r\rightarrow 0\right)\rightarrow 1$, i.e., to look like the 
Fourier transform of a mass conserving smoothing kernel).  
The change in bias due to this term at observed scale 
$k$ is quite simply proportional to the variance on scale $k$, as defined by
the weighting function $I_R\left(r\right)$.   

A similar procedure must be followed with the second $s^2$ term, i.e., 
\begin{equation}
c_{s^2}~P\left(k\right)~
\int \frac{d^3\vq}{\left(2 \pi\right)^3}P\left(q\right)
\FtwoS\left(-\vq,\vk\right)
\ssqfac\left(\vq,\vk-\vq\right)
~\stackrel{k \rightarrow 0}{\longrightarrow}~
\frac{68}{63}~ c_{s^2}~\sigma^2~ P(k) ~.
\end{equation}
All of the other terms go to zero for small $k$.
As in \cite{2006PhRvD..74j3512M}, we now define the observable, renormalized,
linear bias as the sum of bias-like terms 
\begin{equation}
b_\delta = c_\delta+ \left(
\frac{34}{21} c_{\delta^2}+
\frac{68}{63} c_{s^2}+
\frac{1}{2}c_{\delta^3}+\frac{1}{3}c_{\delta s^2}
-\frac{16}{63}c_{st}\right)\sigma^2+
\frac{1}{2}c_{\delta \epsilon^2}~\sigma_\epsilon^2
~.
\end{equation}
Note that this is the only appearance of the parameters 
$c_{\delta^3}$, $c_{\delta s^2}$, and $c_{\delta \epsilon^2}$, 
so they are no longer needed. In fact, the random noise variable $\epsilon$
has completely disappeared, just like it would have if it was only included as
a single term at the end of the Taylor series.

The $P_{mg}\left(k\right)$ result simplifies even more when 
we find, somewhat surprisingly, that the three terms proportional to 
$P\left(k\right)$ in Eq. (\ref{eqmassgal}) are exactly proportional to each 
other, after renormalization and angle-integration.
This means that we can define one merged term that accounts for all of them,
i.e., 
\begin{eqnarray}
c_3~ \sigma^2_3\left(k\right)~P\left(k\right) &\equiv& 
2~ c_{s^2}~P\left(k\right)~
\int \frac{d^3\vq}{\left(2 \pi\right)^3}P\left(q\right)\left[
\FtwoS\left(-\vq,\vk\right)
\ssqfac\left(\vq,\vk-\vq\right)-\frac{34}{63}\right]
\\ \nonumber &+&
2~ c_{\tttheta}~P\left(k\right)~ \int \frac{d^3\vq}{\left(2 \pi\right)^3}
P\left(q\right)
\left[\frac{3}{2}D_S^{\left(3\right)}\left(\vq,-\vq,-\vk\right)-2~
\FtwoS\left(-\vq,\vk\right)
D_S^{\left(2\right)}\left(\vq,\vk-\vq\right)\right]
\\ \nonumber &+&
2~ c_{st}~P\left(k\right)~ \int \frac{d^3\vq}{\left(2 \pi\right)^3}
P\left(q\right)
\left[
D_S^{\left(2\right)}\left(-\vq,\vk\right)
\ssqfac\left(\vq,\vk-\vq\right)+\frac{8}{63}\right] \\ \nonumber
&=&
\frac{32}{105}\left(c_{st}-\frac{5}{2}c_{s^2}+\frac{16}{21}c_{\tttheta}\right) 
\sigma_3^2\left(k\right) ~P\left(k\right) ~, 
\end{eqnarray}
where
\begin{equation}
\sigma^2_{3}\left(k\right)\equiv 
\int d\ln r~ \Delta^2\left(k r\right)~ I_R\left(r\right)~.
\end{equation}
Note that the inclusion of the $s^2$ term in this redefinition is convenient
but not at all necessary, because it is perfectly well-behaved, and the
redefinition does not remove all appearances of 
the parameter $c_{s^2}$. The reason to include this term in the 
redefinition is that, presumably, a fit to data using $c_{s^2}$ and $c_3$ will
show less degeneracy between the two parameters if the functions they multiply
do not have substantial components which have identical form.  

Finally, we define normalized parameters $\tdtwo=
c_{\delta^2}/b_\delta$,
$\tstwo=c_{s^2}/b_\delta$, 
and $\tbthree=c_{3}/b_\delta$ to produce
the power spectrum
\begin{equation}
P_{mg}(k)=
b_\delta \left(P_{\rm NL}(k)+
\tbthree~ \sigma^2_{3}\left(k\right) P\left(k\right)
 + \int \frac{d^3\vq}{\left(2 \pi\right)^3}P\left(q\right)
P\left(\left|\vk-\vq\right|\right)
\FtwoS\left(\vq,\vk-\vq\right)\left[\tdtwo+
\tstwo \ssqfac\left(\vq,\vk-\vq\right) \right]
\right)~.
\label{eqfinalPmg}
\end{equation} 

\begin{figure}
\resizebox{\textwidth}{!}{\includegraphics{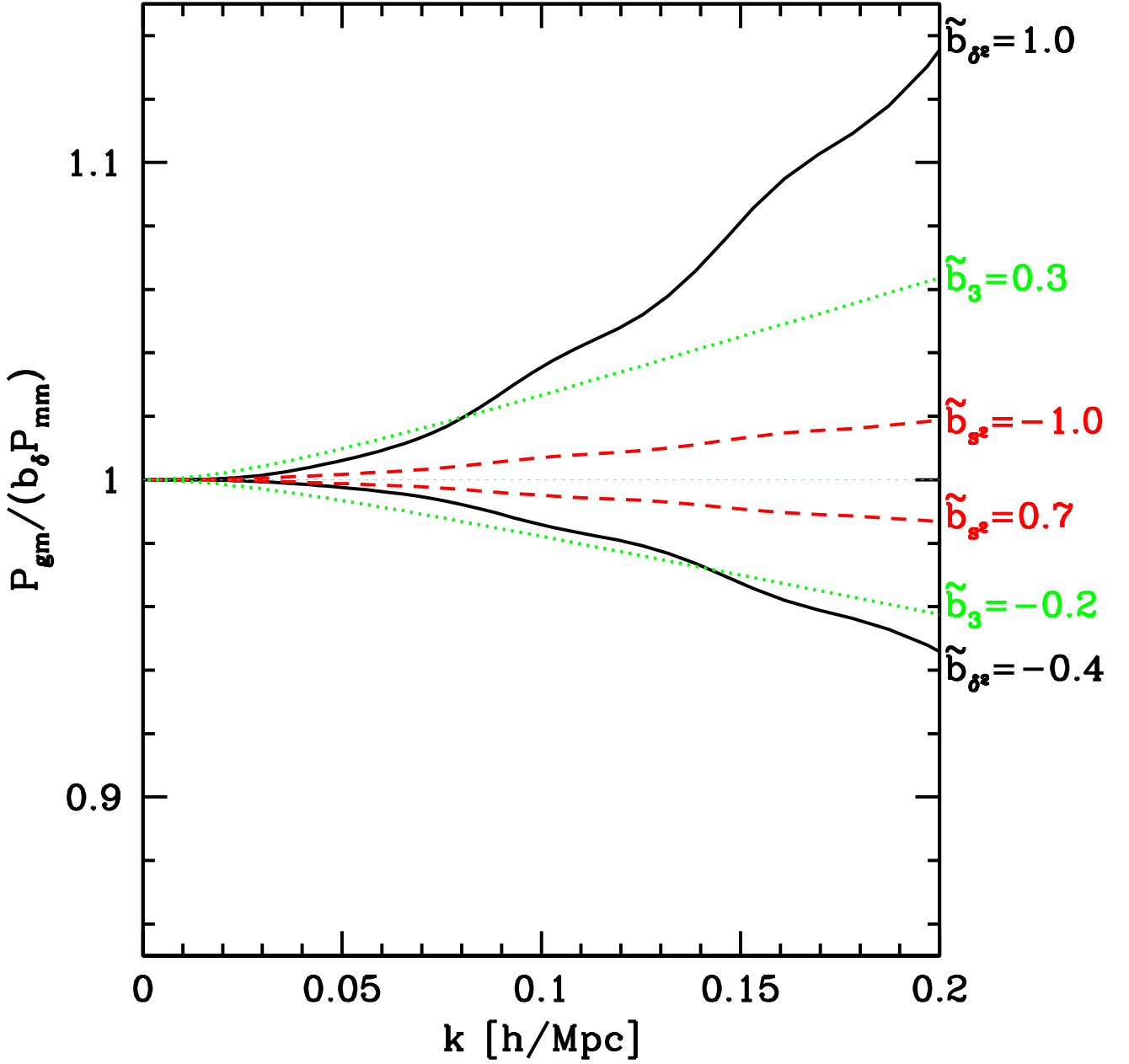}}
\caption{
Bias terms in Eq. (\ref{eqfinalPmg}), for the galaxy-mass cross-power spectrum,
at $z=1$.  
The black (solid) line shows the term 
proportional to $\tdtwo$, red (dashed) shows
$\tstwo$, and green (dotted) shows  $\tbthree$.
The coefficient values are chosen to match those in the more important
galaxy-galaxy power spectrum shown in Fig. \ref{figbasicgg}.
}
\label{figbasicmg}
\end{figure}

The final expression has two new terms relative to the version from the 
$\delta$-only Taylor series in 
\cite{2006PhRvD..74j3512M}.  The term associated with $\sigma^2_{3}$ is
more like a true $k$-dependent bias, in the sense that the power at a given $k$ 
is still proportional to the matter power spectrum at that $k$, just multiplied
by a $k$-dependent factor; while the other term, associated with 
$\tstwo$, mixes power from a range of scales. These terms come from the 
correlation of the linear and second order parts, respectively, of the mass 
density field with the galaxy field.
Figure \ref{figbasicmg} shows the effect of all the terms, for a typical 
$\Lambda$CDM model, at $z=1$.
We see that the $\tstwo$ term is actually quite small relative to the 
others, for similar values of the bias parameters.
In this paper the parameter values are completely arbitrary, 
simply chosen to make the
different effects comparable in size in the more easily observable
galaxy-galaxy power spectrum, $P_{gg}$, where the effect of the $\tstwo$
term is substantially larger (Fig. \ref{figbasicgg}).
The $\tbthree$ term, on the other hand, can have a larger effect on 
$P_{mg}$, relative to its effect on $P_{gg}$. 

Note that
we could have, completely equivalently, left $\ttheta$ as our independent variable
while redefining $s^2$ to make it non-zero only at 3rd order.  All differences
in the resulting equations would be numerical factors which can
be removed by redefining the parameters.  The least trivial looking of these
changes would be changing
$\tstwo \ssqfac\left(\vq,\vk-\vq\right)$ term in 
Eq. (\ref{eqfinalPmg}) to 
$\tilde{b}_{\ttheta}D_S^{\left(2\right)}\left(\vq,\vk-\vq\right)$; however, 
the simple relation between $D_S^{\left(2\right)}$ and 
$\ssqfac$ (Eq. \ref{eqDFrelation}) means that this change is
equivalent to redefining $\tstwo$ and $\tdtwo$.

\subsubsection{Galaxy-Galaxy power spectrum}

We now compute the cross-power spectrum between two types of galaxies, each 
with a set of bias parameters represented by the letters $a$ and $b$.  The 
power spectrum of a single type of galaxy is of course obtained by taking 
equal bias parameters for each type.
\begin{eqnarray}
\label{eqfirstPab}
P_{ab}\left(k\right)&=&
a_\delta b_\delta \left(
P_{\rm NL}(k)+
\left[\tabthree+\tbthree\right]~ \sigma^2_3\left(k\right)~P\left(k\right) 
\right.
\\ \nonumber &+&
\left.
\int \frac{d^3\vq}{\left(2 \pi\right)^3}P\left(q\right)
P\left(\left|\vk-\vq\right|\right)
\FtwoS\left(\vq,\vk-\vq\right)\left[\tadtwo+
\tdtwo+
\left(\tastwo+\tstwo\right)
\ssqfac\left(\vq,\vk-\vq\right) \right]
\right. \\ \nonumber 
&+&  \left.
\frac{1}{2}\int \frac{d^3\vq}{\left(2 \pi\right)^3}P\left(q\right)
P\left(\left|\vk-\vq\right|\right)
\left[\tadtwo \tdtwo+
\left(\tastwo \tdtwo+\tadtwo \tstwo\right)
\ssqfac\left(\vq,\vk-\vq\right) +
\tastwo \tstwo \ssqfac\left(\vq,\vk-\vq\right)^2 
\right]\right) 
\\ \nonumber 
&+&a_\epsilon b_\epsilon\left[ 1 +
\left(\tilde{a}_{\delta^2\epsilon}+\tilde{b}_{\delta^2\epsilon}
\right) \frac{\sigma^2}{2} +
\left(\tilde{a}_{s^2\epsilon}+\tilde{b}_{s^2\epsilon}\right)
\frac{\sigma^2}{3} +
\frac{1}{2}\left(\tilde{a}_{\epsilon^3}\sigma_{\epsilon a^2}^2+
\tilde{b}_{\epsilon^3}\sigma_{\epsilon b^2}^2\right)+
\tilde{a}_{\delta\epsilon}\tilde{b}_{\delta\epsilon} 
\sigma^2+
\tilde{a}_{\epsilon^2}\tilde{b}_{\epsilon^2} 
\frac{\sigma_{\epsilon ab}^2}{2} \right]P^{\epsilon}_{ab}
 ~.
\end{eqnarray} 
The first two lines in Eq. (\ref{eqfirstPab}) are the terms proportional 
to the linear bias factor of one type of galaxy or the other, and are thus 
essentially just the $P_{mg}$ result re-written (including already all of the
same renormalizations).  The third line contains the new terms due to 
cross-products of the 2nd order bias factors.
The last line contains cross-terms involving the random variables 
$\epsilon_a$ and $\epsilon_b$, which we have taken to be 
possibly locally correlated with cross-power spectrum 
$P_{ab}^\epsilon$, and cross-variance 
$\sigma^2_{\epsilon ab}\equiv \left<\epsilon_a \epsilon_b\right>$. 

In the $k\rightarrow 0$ limit the new terms in the third line of 
Eq. (\ref{eqfirstPab}) are not zero, but are 
$k$-independent, i.e., they look like locally correlated white noise:
\begin{eqnarray}
\label{eqstochasticity}
\frac{1}{2}& &\int \frac{d^3\vq}{\left(2 \pi\right)^3}P\left(q\right)
P\left(\left|\vk-\vq\right|\right)
\left[\tadtwo \tdtwo+
\left(\tastwo \tdtwo+\tadtwo \tstwo \right)
\ssqfac\left(\vq,\vk-\vq\right) +
\tastwo \tstwo \ssqfac\left(\vq,\vk-\vq\right)^2 
\right] \\ \nonumber & &
~\stackrel{k \rightarrow 0}{\longrightarrow}~ 
\frac{1}{2} 
\left(\tadtwo+\frac{2}{3}\tastwo \right)
\left(\tdtwo+\frac{2}{3}\tstwo \right)
\int \frac{d^3\vq}{\left(2 \pi\right)^3}P\left(q\right)^2 ~.
\end{eqnarray}
It is interesting to note that these shot-noise-like terms in the power 
spectrum come from the same terms in the original galaxy density Taylor series 
which produced a non-zero contribution to the mean density.  This is consistent
with our expectation that white noise must be associated with
non-conservation of the field.
As in \cite{2006PhRvD..74j3512M}, we can absorb these constant terms into the 
observable
noise matrix, but first we need to discuss the $\epsilon$-related terms.

We define the lowest order $\epsilon$-related term in the last line of 
Eq. (\ref{eqfirstPab}) 
to be $N_{0ab}=a_\epsilon b_\epsilon 
P_{ab}^\epsilon$. If we were only calculating to lowest order, this would 
be the usual galaxy shot-noise. 
The rest of the terms are also constants ($k$-independent), so they can be
simply interpreted as renormalizing this
noise matrix, i.e., in spite of the apparent large number of new terms, there
is actually nothing new here at all.
After renormalization, the result is a completely 
general effective noise matrix for the galaxies, i.e., some choice of the
bias parameters can produce any mathematically legitimate matrix. 
Altogether, the formal redefinition is:
\begin{eqnarray}
\label{eqNrenorm}
N_{ab} &=& N_{0ab}\left[ 1 +
\left(\tilde{a}_{\delta^2\epsilon}+\tilde{b}_{\delta^2\epsilon}
\right) \frac{\sigma^2}{2} +
\left(\tilde{a}_{s^2\epsilon}+\tilde{b}_{s^2\epsilon}\right)
\frac{\sigma^2}{3} +
\frac{1}{2}\left(\tilde{a}_{\epsilon^3}\sigma_{\epsilon a^2}^2+
\tilde{b}_{\epsilon^3}\sigma_{\epsilon b^2}^2\right)+
\tilde{a}_{\delta\epsilon}\tilde{b}_{\delta\epsilon}
\sigma^2+
\tilde{a}_{\epsilon^2}\tilde{b}_{\epsilon^2}
\frac{\sigma_{\epsilon ab}^2}{2} \right] \\ \nonumber
& &+\frac{1}{2}
\left(a_{\delta^2}+\frac{2}{3}a_{s^2} \right)
\left(b_{\delta^2}+\frac{2}{3}b_{s^2}  \right)
\int \frac{d^3\vq}{\left(2 \pi\right)^3}P\left(q\right)^2~.
\end{eqnarray}
The result that we should have a general free noise matrix is 
insensitive to assumptions about the form of the
matrix $P_{ab}^\epsilon$ -- we could start by assuming that
$\epsilon_a$ and $\epsilon_b$ are perfectly correlated (i.e., there is really 
only one random variable), or perfectly independent, and in either case the
renormalizations would generate the extra freedom. 
We do require some intrinsic randomness, i.e., we cannot start with
$P_{ab}^\epsilon=0$ and rely entirely on the noise matrix 
generated by the
density fluctuations (if we want to allow for different types of galaxies to be
uncorrelated, or correlated in a way different from that given by the right-hand
side of Eq. \ref{eqstochasticity}). This is somewhat unsatisfactory as the 
randomness in the initial density field must ultimately be the source of 
randomness in the outcome -- we speculate that higher order density field terms 
will produce
a general noise matrix, so that eventually there will be no need to give 
$\epsilon$ a seed variance.
Note that one should not think too hard about where a noise matrix that is 
nearly diagonal with elements equal to the inverse mean number density of 
galaxies ($\bar{n}_g^{-1}$) comes from
in this picture (aside from observing that it is possible). 
The terms that appear on the right hand side of 
Eq. (\ref{eqNrenorm}) do not need to
add up to the observable noise in any literal sense, because the 
observable noise will contain other, possibly even larger, terms at higher 
order. Eq. (\ref{eqNrenorm}) just shows why it is legitimate to drop the 
undesirable terms (that are non-zero as $k\rightarrow 0$, including all of the
$\epsilon$-related terms) 
in the PT calculation, i.e., because they are redundant with
a free noise matrix. 
One should remember that the idea of Poisson sampling, i.e., the 
$\bar{n}_g^{-1}$ model 
for noise power, was never more than an apparently quite accurate guess --
\cite{2007PhRvD..75f3512S}, for example, found deviations for dark matter 
halos. 

We are left with the final power spectrum:
\begin{eqnarray}
\label{eqfinalPab}
P_{ab}\left(k\right)&=&
a_\delta b_\delta \left(P_{\rm NL}(k)+
\left[\tabthree+\tbthree\right]~ \sigma^2_3\left(k\right)~P\left(k\right)
\right.
\\ \nonumber &+&
\left.
\int \frac{d^3\vq}{\left(2 \pi\right)^3}P\left(q\right)
P\left(\left|\vk-\vq\right|\right)
\FtwoS\left(\vq,\vk-\vq\right)\left[\tadtwo+
\tdtwo+
\left(\tastwo+\tstwo\right)
\ssqfac\left(\vq,\vk-\vq\right) \right]
\right. \\ \nonumber 
&+&  \left.
\frac{1}{2}\int \frac{d^3\vq}{\left(2 \pi\right)^3}P\left(q\right)
\left[
\tadtwo \tdtwo
\left[P\left(\left|\vk-\vq\right|\right)-P\left(q\right)\right] +
\left(\tastwo \tdtwo+\tadtwo \tstwo \right)
\left[\ssqfac\left(\vq,\vk-\vq\right)~
P\left(\left|\vk-\vq\right|\right)-\frac{2}{3}P\left(q\right)\right]
\right.\right. \\ \nonumber & & + \left. \left.
\tastwo \tstwo \left[\ssqfac\left(\vq,\vk-\vq\right)^2
P\left(\left|\vk-\vq\right|\right) -\frac{4}{9} P\left(q\right)\right]
\right]\right) 
\\ \nonumber 
&+&N_{ab} ~.
\end{eqnarray} 
This equation is not as complicated as it may look, including only a few
simple
building blocks:  $P\left(q\right)$, $P\left(\left|\vk-\vq\right|\right)$,
$\FtwoS\left(\vq,\vk-\vq\right)$, 
$\ssqfac\left(\vq,\vk-\vq\right)$, and $I_R(r)$ 
(in $\sigma^2_3\left(k\right)$).

\begin{figure}
\resizebox{\textwidth}{!}{\includegraphics{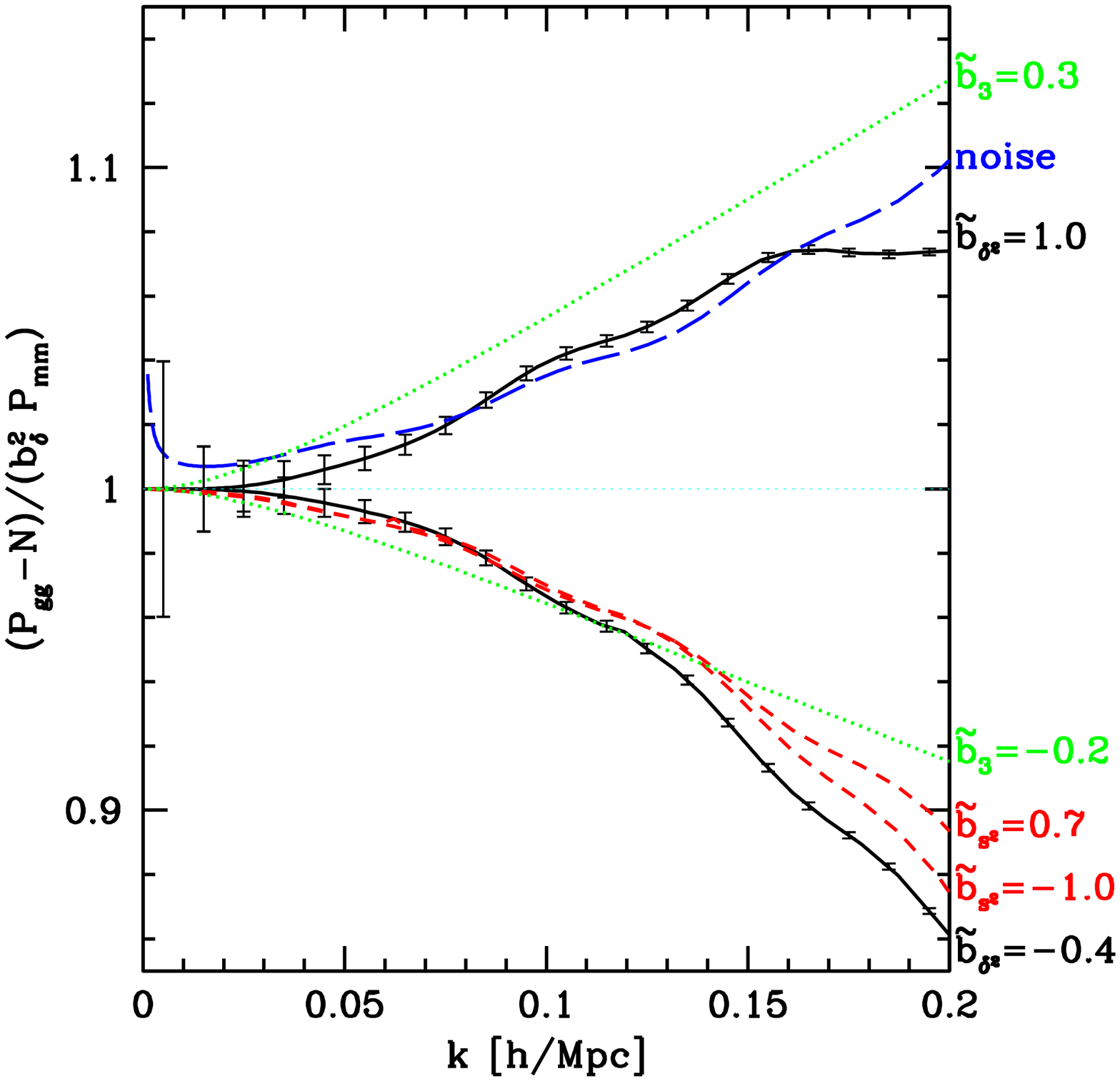}}
\caption{
Effect of various kinds of bias on the auto-power spectrum of a single type of 
galaxy (Eq. \ref{eqfinalPab}), at $z=1$.  The black (solid) line shows the term 
proportional to $\tdtwo$, red (dashed) shows 
$\tstwo$, and green (dotted) shows $\tbthree$, with values of the 
coefficients
labeling the curves (all of the other coefficients are zero in each case).
The blue (long-dashed) line shows the effect of $N$ (white noise), when 
similarly normalized by the mass power spectrum. 
The coefficient values are largely arbitrary, i.e., the lines are only
intended to show the shape of the effect, not to imply anything about the 
magnitude.  The error bars show approximate fractional errors 
on band power measurements from a 100 cubic Gpc/h survey (e.g., 
$\sim 3/4$ of the sky at $1<z<2$). 
}
\label{figbasicgg}
\end{figure}

Figure \ref{figbasicgg} shows examples of the auto-power spectrum for a single 
type of galaxy.  
We see that the effects of each term are somewhat different.  The 
$\tbthree$ term has a greater influence at larger relative to smaller 
scales than the $\tdtwo$ term.  Those two terms can have either
sign, but the $\tstwo$ term is essentially always negative. 
Note that the power spectrum is not linear in the bias parameters, so the
outcome when all of the parameters are varied is more complex than a simple 
sum of the examples we show.
The increase due to the $\tdtwo$ term actually reaches a maximum 
(for $k=0.2\ihmpc$) at $\tdtwo \simeq 0.6$, before declining again as the 
negative
quadratic part comes to dominate (this transition is apparent as the flattening
at the high $k$ end in the figure).

\subsubsection{Bispectrum}

The bispectrum is the three point correlation function 
\cite{2007MNRAS.378.1196K,2005MNRAS.364..620G,2005MNRAS.362.1363P,
2005MNRAS.361..824G,1994ApJ...425..392F} 
in Fourier space. It vanishes if
the density fluctuations are Gaussian.
The bispectrum can be used to measure non-Gaussianity in the primordial 
density  distribution, if any, and non-Gaussianity 
induced by non-linear gravitational evolution and bias
\cite{2008PhRvD..78b3523S,2007PASJ...59...93N,2000ApJ...544..597S,
1998MNRAS.300..747V,1997MNRAS.290..651M}.
\cite{2006PhRvD..74b3522S,2005PhRvD..71f3001S} show that the bispectrum 
is a very powerful addition
to the power spectrum for general cosmological parameter constraints, 
especially on the primordial power spectrum amplitude and slope.
In this section we show the form of the galaxy bispectrum 
in our generalized bias model. 
Only 2nd order terms in the density perturbations
are needed to construct the bispectrum to 4th order.  By definition
bispectrum takes the following form 
\begin{equation}
\langle\delta\left(\textbf{k}_1\right)\delta\left(\textbf{k}_2\right)\delta\left(\textbf{k}_3\right)\rangle=\left(2\pi\right)^3\delta^D\left(\textbf{k}_1+\textbf{k}_2 +\textbf{k}_3  \right)B\left(k_1,k_2,k_3\right),
\end{equation}
where $\delta^D\left(\textbf{k}_1+\textbf{k}_2 +\textbf{k}_3  \right)$ 
means that only closed triangular configurations are non-zero.
In our calculations, we assume that the primordial density fluctuations  
did not have any signature of non-Gaussianity.
The galaxy bispectrum is then 
\begin{eqnarray}
B_g(k_1,k_2,k_3)&=&b_{\delta}^3~ P\left(k_1\right)P\left(k_2\right)
\left[2~\FtwoS\left(\textbf{k}_1,\textbf{k}_2\right)+\tdtwo
+\tstwo \ssqfac\left(\textbf{k}_1,\textbf{k}_2\right) \right]+
2~\hat{b}_{\delta\epsilon} N~b_\delta^2 P\left(k_1\right)+
\hat{b}_{\epsilon^2} N^2
\nonumber\\
 & & + {\rm cyclic~ permutations~ of}~k_1,~k_2,~k_3~,
\end{eqnarray}
where we note that the angle between any two of the $\vk$ vectors is determined
by the length of the third. We have defined 
$\hat{b}_{\delta\epsilon}=\frac{b_{\delta\epsilon}}{b_\delta b_\epsilon}$ and
$\hat{b}_{\epsilon^2}=\frac{b_{\epsilon^2}}{b_\epsilon^2}$, and $N$ is the 
noise power. Here we see directly the convergence between Eulerian and 
Lagrangian bias that we were hoping for -- the new $s^2$ term introduces the
extra configuration dependence in the bispectrum found for Lagrangian bias by 
\cite{2000MNRAS.318L..39C,2002PhR...367....1B,1998MNRAS.297..692C}.
Note that \cite{2001PhRvL..86.1434F} actually compare Lagrangian vs.
traditional (density-only) Eulerian bias in fits to the PSCz bispectrum, but 
did not have
enough statistical power to distinguish them (Eulerian bias was slightly 
preferred).

We see now the purpose in the introduction of the full structure of 
$\epsilon$-related terms. These terms have 
produced exactly the structure needed to correctly represent Poisson noise in 
the 
bispectrum. If the galaxies were a Poisson sampling of the 
underlying biased density field, we would have 
$\hat{b}_{\delta\epsilon}=\frac{1}{2}$
and $\hat{b}_{\epsilon^2}=\frac{1}{3}$ 
\cite{1980lssu.book.....P,2008PhRvD..78b3523S}. 
Even the appearance of the
extra new free parameters, $\hat{b}_{\delta\epsilon}$ and 
$\hat{b}_{\epsilon^2}$,
is necessary, as \cite{2008PhRvD..78b3523S} showed that galaxies in
simple halo models do not obey Poisson sampling exactly, but 
instead follow the more general form we find here, with the values of  
$\hat{b}_{\delta\epsilon}$ and $\hat{b}_{\epsilon^2}$ depending on the details
of the model (in fact, our introduction of this treatment of noise was 
entirely motivated by \cite{2008PhRvD..78b3523S}).

A reduced bispectrum, which does not depend on the mass power spectrum 
amplitude, is often written as
\begin{equation}
Q_X\left(k_1,k_2,k_3\right)=\frac{B_X\left(k_1,k_2,k_3\right)}{
P_X\left(k_1\right)P_X\left(k_2\right)+P_X\left(k_2\right)P_X\left(k_3\right)+
P_X\left(k_3\right)P_X\left(k_1\right)}~.
\end{equation}
The reduced galaxy bispectrum, to leading order, is then,
\begin{equation}
Q_g\left(k_1,k_2,k_3\right)=b_\delta^{-1}\left[Q_m\left(k_1,k_2,k_3\right)+
\tdtwo+\tstwo
\frac{P\left(k_1\right)P\left(k_2\right)
\ssqfac\left(\textbf{k}_1,\textbf{k}_2\right)+{\rm cyclic ~
perms.}}{
P\left(k_1\right)P\left(k_2\right)+P\left(k_2\right)P\left(k_3\right)+
P\left(k_3\right)P\left(k_1\right)}\right]~,
\label{eqQg}
\end{equation}
where $Q_m$ is the reduced bispectrum of the mass density perturbations.
The noise terms, which we have dropped from this presentation of $Q_g$, 
undermine the elegance of using $Q_g$. We suspect that it will be
more straightforward
to interpret noisy observations using a simultaneous fit to $P_g$ and $B_g$, 
rather than going through $Q_g$. 

\begin{figure}
%\resizebox{\textwidth}{!}{\includegraphics{bis.1.eps}}
\subfigure{\includegraphics[width=0.49\textwidth]{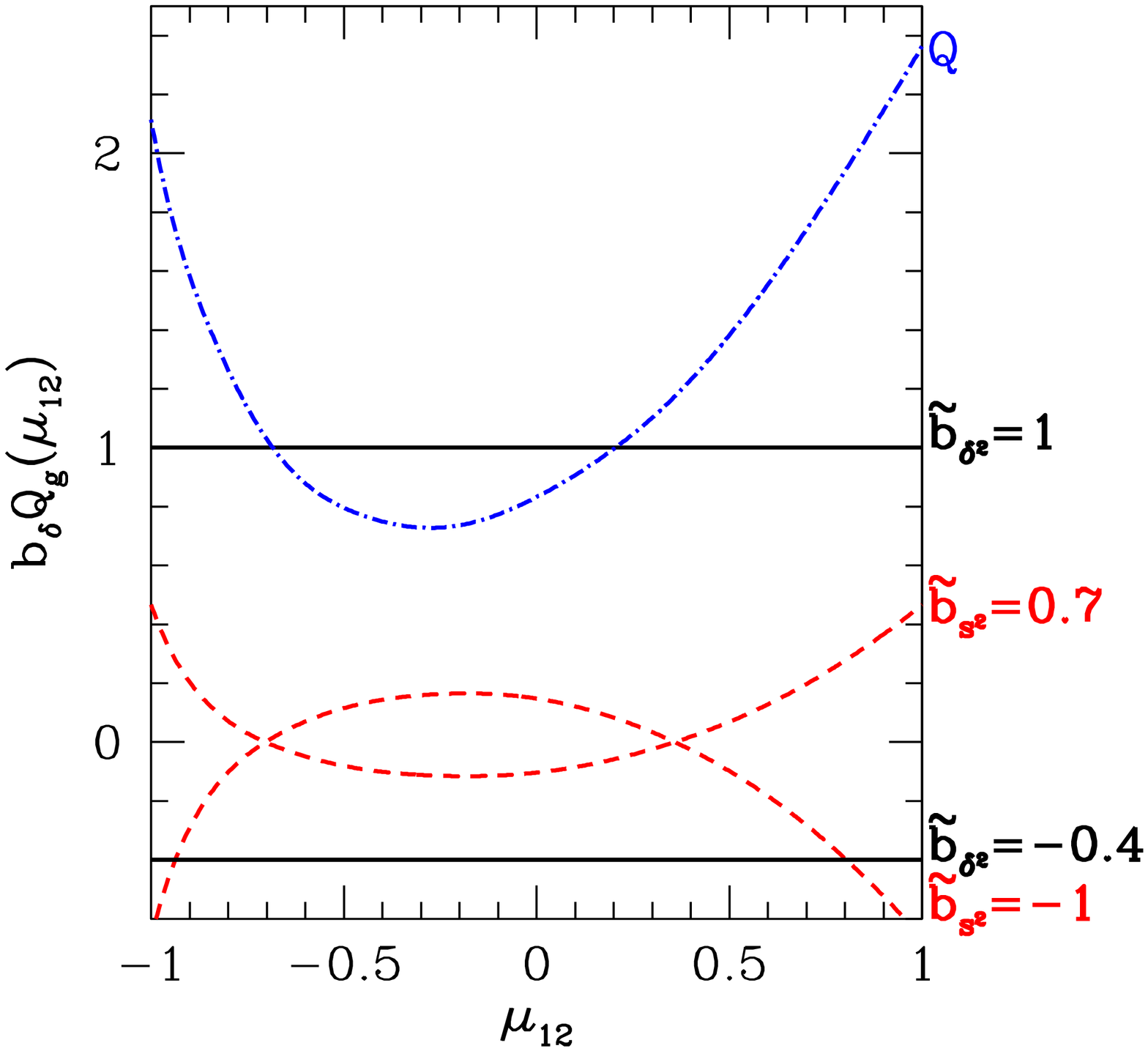}}
\subfigure{\includegraphics[width=0.49\textwidth]{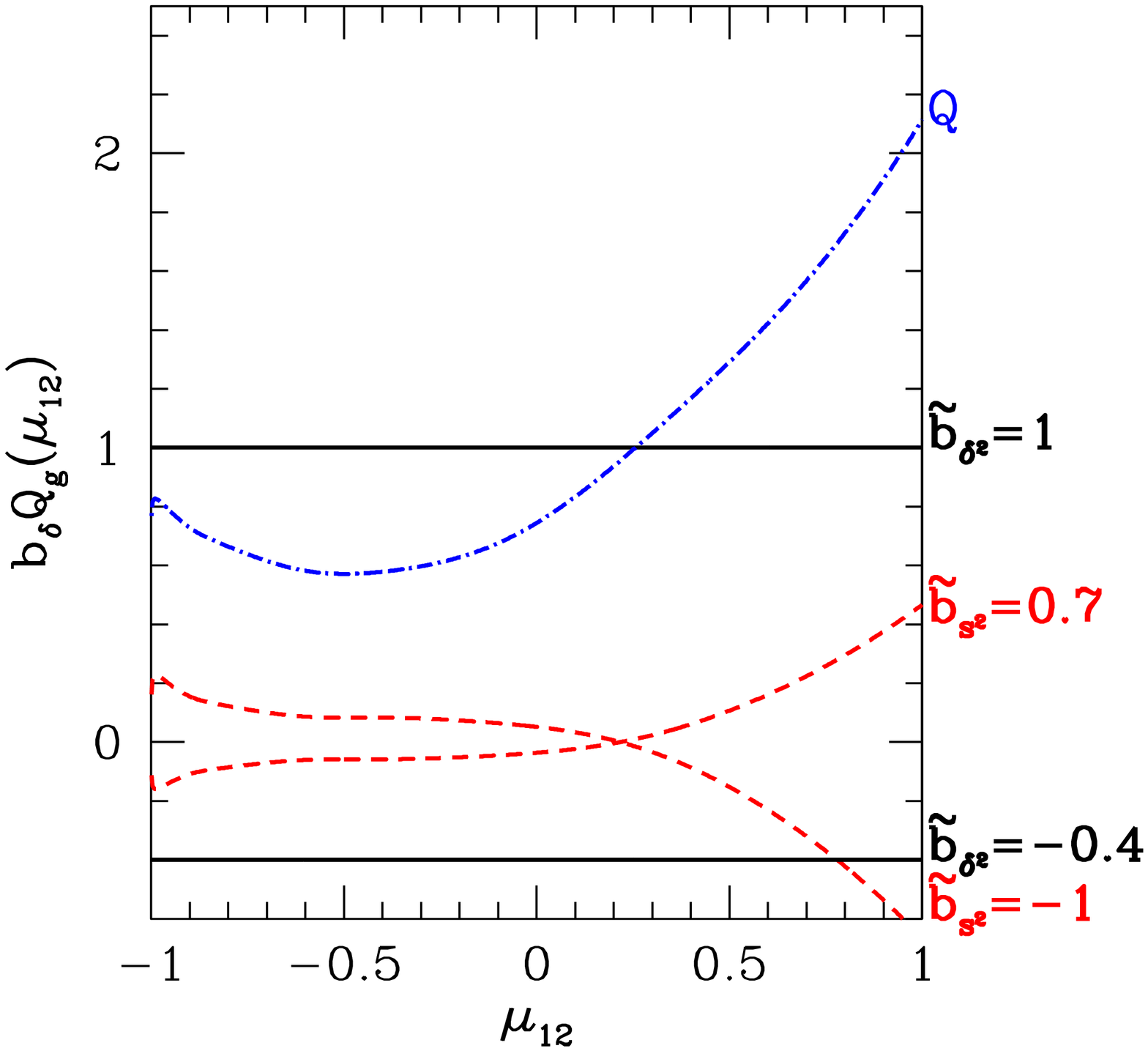}}
\caption{Quantities contributing to the
reduced bispectrum, $Q_g$ (Eq. \ref{eqQg}), as a function of 
$\mu_{12}=\vk_1\cdot\vk_2/k_1 k_2$. 
The left panel shows
$k_1=0.1 \ihmpc$, $k_2=0.2 \ihmpc$, while the right shows $k_1=k_2=0.1\ihmpc$
(in this case, recall that $\vk_3=-\left(\vk_1+\vk_2\right)$, so $k_3=0$ when 
$\mu_{12}=-1$).  The blue, dot-dashed,
curve shows the mass bispectrum $Q_m$, the black, solid, lines represent
$\tilde{b}_{\delta^2}$, and the red, dashed, curves are the new 
$\tilde{b}_{s^2}$ term 
(new to Eulerian bias models, although already present in Lagrangian bias 
models \cite{2000MNRAS.318L..39C,2002PhR...367....1B,1998MNRAS.297..692C}). 
The parameter values were chosen to match the power
spectrum figures.
} 
\label{figbispectrum}
\end{figure}

Figure \ref{figbispectrum} shows some examples of the reduced bispectrum and
bias terms.
$Q_g$ has been discussed as a means to measure $b_\delta$, because, unlike
$P_g$, it is only sensitive to $b_\delta$, not to the 
amplitude of the mass power spectrum. It has always been necessary to
marginalize over $\tdtwo$ in this approach
\cite{2002MNRAS.335..432V,2001PhRvL..86.1434F}, and now we have an extra 
possibility, degeneracy with $\tstwo$. It is still possible to measure all 
the parameters independently, because $\FtwoS$ and
$S$ differ by more than an additive constant; however, it 
would be helpful if a plausible upper limit on $\tstwo$ could be determined 
using simulations. 

Other higher order statistics, like Fourier phase statistics 
\cite{2005PASJ...57..709H}, the
trispectrum \cite{2005PhRvD..71f3001S}, or the probability distribution
function of counts in cells \cite{2004ApJ...602...26S},
could also be considered.

\section{Miscellaneous further extensions \label{secextensions}}

In this section we discuss a few further extensions of the baseline approach 
to bias outlined in the previous section. 
In \S \ref{secnonlocal} we discuss additional 
short-range non-locality in the bias relation. In \S \ref{secredshift}
we discuss briefly the new considerations that arise when one goes to redshift
space. Finally, in \S \ref{secnonG} we discuss non-Gaussian initial conditions.

\subsection{Short-range non-locality \label{secnonlocal}}

So far, our model has included non-local dependence of the galaxy density on
the mass density, but only in the form of local dependence on 
$\nabla_i \nabla_j \phi$ and $\nabla_i v_j$, which are in turn determined by 
the density field through gravitational evolution.
For completeness, we now consider relatively short-range non-locality that
might be caused by hydrodynamics or the highly non-linear details of galaxy 
formation, i.e.,
\begin{equation}
\delta_g\left(\vx\right)=f\left[\delta\left(\vx^\prime\right)\right]~,
\end{equation}
where the galaxy density at $\vx$ depends on the mass density at all points
$\vx^\prime$ roughly 
obeying $\left|\vx-\vx^\prime\right|\lesssim R$. We assume that
$R$ is small in the sense that $k^2 R^2<<1$, where $k$ is the observed 
wavenumber.
First, we expand $\delta_g$ as a Taylor series in $\delta$, i.e.,
\begin{equation}
\delta_g\left(\vx\right)=f\left[\delta\left(\vx^\prime\right)\right]
=f\left[0\right]+
\int d\vx^\prime~ K\left(\left|\vx-\vx^\prime\right|\right)
\delta\left(\vx^\prime\right)+...~,
\label{eqnonlocalexpanddelta}
\end{equation}
where 
$K\left(\left|\vx-\vx^\prime\right|\right)$ is the kernel of derivatives of
galaxy density at $\vx$ with respect to mass density at $\vx^\prime$.
We allow an almost arbitrary form for $K$, 
except that it must fall
to zero outside a typical scale $R$, and it must be isotropic.
We now shift the integration variable to $\Delta \vx = \vx-\vx^\prime$, and
Taylor expand in $\Delta \vx$, i.e., taking only the linear term,
\begin{eqnarray}
\delta_g\left(\vx\right)&=&
\int d\Delta\vx~K\left(\left|\Delta \vx\right|\right)
\delta\left(\vx+\Delta \vx \right)=
\int d\Delta\vx~ K\left(\left|\Delta \vx\right|\right)\left[
\delta\left(\vx\right)+\frac{d\delta}{dx_i}\left(\vx\right)\Delta x_i+
\frac{1}{2}\frac{d^2\delta}{dx_i dx_j}\left(\vx\right)\Delta x_i\Delta x_j
+...\right] \\ \nonumber
&=&\delta\left(\vx\right)
\int d\Delta\vx~ K\left(\left|\Delta \vx\right|\right)+
\frac{d\delta\left(\vx\right)}{dx_i}
\int d\Delta\vx~ K\left(\left|\Delta \vx\right|\right)
\Delta x_i+
\frac{1}{2}\frac{d^2\delta\left(\vx\right)}{dx_i dx_j}
\int d\Delta\vx~ K\left(\left|\Delta \vx\right|\right)
\Delta x_i\Delta x_j
+...
\end{eqnarray}
(this derivation was inspired by \cite{2007arXiv0704.3205N}).
The simple integral over $K$ in the first term is naturally defined to be the
standard linear bias, $b_\delta$. 
The 2nd term, integrating $K~\Delta x_i$, must be
zero by the symmetry of the kernel. The third term, integrating 
$K~ \Delta x_i \Delta x_j$ must be zero by symmetry if $i\neq j$, but if 
$i=j$, the integral for a  generic kernel will give a result of order $R^2$ 
times the simple integral over the kernel in the first term, i.e., 
the integral will give a result of order $\sim b_\delta R^2 \delta^K_{ij}$. 
Therefore, we have for the galaxy density
\begin{equation}
\delta_g\left(\vx\right)=b\left[\delta\left(\vx\right)
+ \frac{\tbR }{2} R^2\nabla^2\delta\left(\vx\right)\right]
+...
\end{equation}
where $\tbR$ is of order unity (e.g., if the kernel was a Gaussian with rms
width $R$, $\tbR$ would be exactly 1). The loophole in the argument that
$\tbR \sim 1$ is if the kernel has substantial positive and negative parts,
and these are tuned to almost perfectly cancel in the average over the 
whole 
kernel, making the average much smaller than the fluctuations. On the other
hand, if the kernel does not have 
significant negative parts, one could even argue that $\tbR$ should be 
not just ${\mathcal O}\left(1\right)$, but also positive.
In Fourier space we have
\begin{equation}
\delta_g\left(\vk\right)=b\left[1
- \frac{\tbR}{2} R^2 k^2 \right]
\delta\left(\vk\right)
+...
\end{equation}
and the power spectrum is
\begin{equation}
P_g\left(k\right)=b^2\left[1
- \tbR~ R^2 k^2 \right]
P\left(k\right)
+...
\label{eqPnonlocal}
\end{equation}
Fits including $\tbR$ can include a prior that $\tbR$ is not much greater
than one, and possibly also positive, although this will only be useful if 
consideration of galaxy formation physics can place an upper limit 
on $R$. Note that, if this program for modeling non-locality is to succeed,
$k^2 R^2$ becomes a second small parameter, in addition to the fluctuation
amplitude, so it is not necessarily necessary to include terms simultaneously 
higher order in both $k^2 R^2$ and $\delta$.

The reader may be tempted at this point to conclude that all we have found is
that short-range non-locality can be modeled by assuming the galaxy density 
simply depends on a smoothed version of the density field -- expanding 
the 
smoothing kernel would generally produce the same $k^2 R^2$ term. The truth
is not quite so simple. If we follow the same procedure on the next, 
${\mathcal O}\left(\delta^2\right)$, term that would appear in 
Eq. (\ref{eqnonlocalexpanddelta}), we find not just the new term that would 
come from
using the square of a smoothed field, $R^2 \delta~\nabla^2 \delta$, but also 
a term $R^2 \left(\vnabla \delta\right)\cdot \left(\vnabla \delta\right)$, with
the two terms generally multiplied by independent bias parameters. Together,
these two terms are equivalent to assuming that the galaxy density depends on
both the square of the smoothed density field and, independently, a smoothed 
version of the square of the un-smoothed density field.
Generally, the correct procedure for representing short-range non-locality 
appears to be to write down all possible 
scalar higher derivative terms, each with its own bias parameter, and a 
factor of $R$ for every derivative. 

Similar arguments can be made for the noise. If it is correlated on
scale $R$, smaller than the scale of observation, one generically expects
the noise power spectrum to look like
\begin{equation}
P_N\left(k\right)=\left[1
- \tilde{N}_R~ R^2 k^2 \right]~N
+...~,
\end{equation}
where $N$ is the usual large scale white noise, and $\tilde{N}_R$ is of order 
unity for generic noise correlation functions. 

\subsection{Redshift Space \label{secredshift}}

Allowing for redshift-space distortions changes the symmetry considerations 
that we used to decide which variables galaxy clustering could depend on.
The radial direction can now be special.  For example,  
$\nabla_\parallel v_\parallel$ and $\nabla_\parallel \nabla_\parallel \phi$
are now allowed in the Taylor series, where $\parallel$ indicates the radial
direction.  The non-locality kernel in \S \ref{secnonlocal} can depend 
separately on the radial coordinate as well, which will lead to an 
$R^2_\parallel k^2_\parallel$ term. All of these terms generally come with an
unknown bias parameter. 
None of these considerations are needed in the usual approach to redshift-space
distortions pioneered for galaxies by \cite{1987MNRAS.227....1K}, because the 
transformation
from real to redshift space is applied to the already biased field, and 
does not involve any new unknown functions. 
The \lyaf\ represents a counter-example 
\cite{2000ApJ...543....1M,2003ApJ...585...34M}, 
where the already redshift-distorted
optical depth field, $\tau$, undergoes the local non-linear transformation
$\exp\left(-\tau\right)$ to produce the observed transmitted flux fraction
field. While the form of this transformation is completely known, it applies
to the un-smoothed optical depth field, which is sufficiently non-linear that
one cannot hope to use the in-this-case-actually-computable Taylor series 
coefficients to describe very large scale clustering -- the observable bias
parameters will inevitably receive perturbatively un-computable contributions 
from higher 
order terms. Consistent with this picture, \cite{2003ApJ...585...34M} showed 
that the standard \cite{1987MNRAS.227....1K} form for the large scale 
redshift-space power spectrum fit the \lyaf\ power spectrum well, as long as 
the distortion parameter $\beta$ was a free parameter, rather than the usual
$\beta = \left(d\ln D/d\ln a\right) b^{-1}$. This is equivalent to introducing
a $\nabla_\parallel v_\parallel$ term with a free bias parameter.
The cautious reader may wonder whether the non-linear 
transformation involved in the usual \cite{1987MNRAS.227....1K} redshift-space
distortion calculation, when taken to higher order as in 
\cite{1998MNRAS.301..797H}, may lead to the same problem of renormalization of
the standard \cite{1987MNRAS.227....1K} form of large-scale power, in a way 
that might
look like velocity bias, for example. We leave this question, and further 
consideration of 
redshift-space distortions in the renormalized bias approach for future work.  

\subsection{Primordial non-Gaussianity \label{secnonG}}

When considering the model for non-Gaussian initial conditions where 
$\phi\left(\vx\right)=\zeta\left(\vx\right) +\fnl \zeta^2\left(\vx\right)$, 
where $\zeta$ is a Gaussian 
variable with the primordial power spectrum that we usually associate with 
$\phi$, \cite{2008PhRvD..77l3514D,2008PhRvD..78l3519M} found the need for a 
bias term directly proportional to 
$\zeta$, which looks for practical purposes like a direct dependence on 
$\phi$ (see also \cite{2008arXiv0808.0044S}).
This may seem inconsistent with the considerations of this paper,
where we excluded any direct dependence on $\phi$. The explanation for this is
that $\zeta$ does not obey the principle that led us to exclude dependence on
$\phi$ -- a homogeneous change in $\zeta$ is observable, essentially as a
change in the primordial power spectrum amplitude, which of course affects 
galaxy 
formation and clustering. This answers the question that was unanswered in 
\cite{2008PhRvD..78l3519M}: whether the term should be considered to be a bias 
parameter 
multiplying $\zeta$ or $\phi$ -- it makes no difference at lowest order, but if
a higher order calculation is needed, the answer clearly is that the dependence
should be on $\zeta$.  

\section{Conclusions \label{secconclusions}}

The central result of this paper is Eq. (\ref{eqfinalPab}), which shows the 
most general
galaxy power spectrum that can be derived starting from expanding the galaxy 
density as a Taylor series in the local values of $\delta$, $\partial_i v_j$, 
$\partial_i \partial_j \phi$. 
This power spectrum depends on only two new
parameters, beyond the usual linear bias, shot-noise, and 2nd order density 
bias.
One of the parameters quantifies 2nd order dependence on the 
magnitude of the 
tidal tensor (or, equivalently after reparameterizations, the difference
between velocity divergence and density), and the other parameter  
multiplies a set of 3rd order terms that collectively appear as a 
$k$-dependent bias 
proportional to the linear variance on scale $k$. 
Eq. (\ref{eqfinalPab}) allows for cross-correlating different types of 
galaxies, each with its own set of bias parameters, but the power spectrum of
a single type of galaxy can be obtained from it by simply setting the bias 
parameters for the two types equal to each other. We also give the 
the cross-spectrum between mass and galaxies explicitly, in Eq.  
(\ref{eqfinalPmg}) (this can of course be
obtained from Eq. \ref{eqfinalPab} by setting the linear bias to 1 and  all of 
the other  bias parameters to zero for one type of galaxy).
In Eq. (\ref{eqQg}) we give the bispectrum of galaxies in this model, which 
includes new dependence on the 2nd order tidal tensor term. Eq. 
(\ref{eqQg}) also shows how 
including a Gaussian white noise variable $\epsilon$ as an expansion variable 
in the original Taylor 
series for galaxy density allows for reproduction of the non-trivial appearance
of Poisson-sampling noise in the bispectrum, or more general non-trivial 
noise properties.
In \S \ref{secnonlocal} we explain how short-range non-locality (from 
hydrodynamics or highly non-linear galaxy formation) can be modeled
as a derivative expansion. 

Since no symmetry prevents it, the galaxy density should have at least some
small dependence
on these new terms -- the question is just how much.
It might have been easy to miss this dependence in past studies
\cite{2008arXiv0805.2632J}, as the $k$
dependence is not enormously different from the density-only model, and 
appears in a range of scales where deviations 
from the density-only model could be interpreted as even higher order effects, 
or confused with shot-noise.
The new terms may not all be necessary, even at the level of future precision
data, but this should be demonstrated, not simply assumed, 
i.e., it would be good if they were all considered and bounded.
To distinguish the terms in simulations, it will be useful to look at the 
mass-galaxy power spectrum, the
galaxy-galaxy power spectrum, and the bispectrum simultaneously (ideally even
the bispectra mixing mass and galaxies, which will be simple to write down,
and higher order statistics).

While one can freely marginalize over the parameters of the extended model 
when interpreting future high precision clustering measurements,
one can also think of this general model as a framework for interpreting 
numerical simulations or other specific models of galaxy/halo formation. 
For a long time, the linear bias parameter has been a 
useful way to condense simulation predictions for very large-scale clustering 
into one number per type of object, rather than simply reporting results 
for free functions $P_g(k)$. 
The parameters of the perturbative model should similarly be a useful way to 
condense perturbative-scale clustering down to a small, well-motivated, 
set of numbers, rather
than discussing scale dependent bias as a free function $b(k)$ (or 
parameterizing it in arbitrary ways \cite{2005MNRAS.362..505C}).
In the most optimistic case, both the halo-based approach and the PT approach
will work very well, and be complementary in that the PT approach
will provide a clear set of large-scale parameters to be calibrated by the 
halo-based approach that includes smaller-scale information.

Some other questions for followup work include:
\begin{itemize}
\item Are there other terms that we should include?
\item The equivalences that led to the need for only a single bias parameter 
at 3rd order should be investigated further.
It seems likely that there are relations like 
$\ttheta_2=\frac{2}{7}s_1^2-\frac{4}{21}\delta_1^2$ which we 
have not taken into account. Note that some of the parameters that are 
unnecessary in the present calculations may become necessary when calculations
are done to higher order.
\item Redshift-space distortions, touched on in \S \ref{secredshift}, should
be computed explicitly within the renormalized bias model.
\item While this property has not been exploited very well in the past, 
the scale where PT breaks down should be internally determinable.
If calculations are pushed to at least one higher order, the breakdown scale
should be evident as the place where the difference between the two highest
orders calculated starts to matter. In the past, PT has acquired a 
reputation for limited accuracy because this kind of testing has not been done,
while the calculations were pushed beyond the point where there was good
reason to expect them to work well. In the future, very high precision, world, 
the primary concern
for PT should not be simple breakdown of the perturbative expansion, but 
instead insufficiently general modeling, e.g., missing terms like the ones in 
this paper. High precision goodness-of-fit tests should also help establish 
reliability.
\item The connection between this approach to bias and renormalization
group/resummation approaches to the non-linear mass clustering 
\cite{2007PhRvD..75d3514M,2008PhRvD..77b3533C,2006PhRvD..73f3519C,
2006PhRvD..73f3520C,2008PhRvD..78j3521B,2008PhRvD..78h3503B,
2008JCAP...10..036P,2008PhRvD..77f3530M,2008ApJ...674..617T,
2008MPLA...23...25M,2007PhRvD..76h3517I,2007JCAP...06..026M}
could be 
considered. \cite{2008arXiv0805.2632J} showed that our approach to bias 
works well when
compared to simulations as long as PT describes the mass power spectrum well
\cite{2006ApJ...651..619J},
but it isn't clear what one should do when standard PT no longer describes the 
mass power well, but more sophisticated methods do. 
\item Time evolution of bias can be considered from the point of view of this
paper \cite{2008PhRvD..77d3527H,2008MNRAS.385L..78P,2005A&A...430..827S,
2001ApJ...550..522B,2000ApJ...537...37T,1999ApJ...510..541T,
1998ApJ...500L..79T,1996ApJ...461L..65F,1995ApJS..101....1M}. 
\item One clear loophole in all of these arguments exists if 
long-range effects of radiation sources
affect galaxy clustering, e.g., through reionization 
\cite{2006ApJ...640....1B,2007JCAP...10..007C,2005MNRAS.360.1471M}
(long enough range to make $k^2 R^2$ not a good expansion parameter). If these
effects are small, some perturbative method can probably be used, but it would
have to be something outside the scope of this paper.  
\item Eventually, one may want to correlate properties of galaxies other than 
density, e.g., ellipticity, galaxy orientation, etc.
\cite{2008MNRAS.389.1266L,2008ApJ...681..798L,2008arXiv0811.1995F,
2008arXiv0809.3790O,2008MNRAS.tmp.1428S,2008arXiv0808.0203S,
2008ApJ...688..742H,2007ApJ...670L...1L,2004ApJ...614L...1L,
2002ApJ...567L.111L,2002MNRAS.332..788M,2002MNRAS.332..325P,
2001MNRAS.320L...7C,2000ApJ...543L.107P}. 
These correlations should be describable by a
similar approach to the one here, except with modified symmetry 
considerations. For example, a traceless tensor observable can be linearly 
related to 
$s_{ij}(\vx)$ by a scalar bias parameter, but not to $\delta(\vx)$.
\end{itemize}

Finally, the background motivation for this work deserves re-emphasis: Fig. 
\ref{figmodes} shows that future redshift surveys will contain orders of 
magnitude more information than present surveys. Fig. \ref{figbasicgg} shows
that there will be a wide range of scales (e.g., very roughly a factor of 4 in 
$k$ or 64 in number of modes), in which corrections to linear theory
will be necessary but still fractionally small, i.e., amenable to a
perturbative treatment, for realistic planned surveys.  
To exploit this information optimally will require rigorous modeling of
clustering, far beyond what has been done in the past.

We thank Roman Scoccimarro for suggesting that we consider dependence on
$\nabla_i \nabla_j \phi$ in addition to $\theta$, and Adam Lidz, 
Neal Dalal, and Latham Boyle for helpful discussions. PM acknowledges support 
of the Beatrice D. Tremaine Fellowship.

\section{Appendix:  PT Basics}

Standard gravitational PT is 
well-described in, e.g., \cite{2002PhR...367....1B}.
Here we list some of the relevant facts that we use.

The density perturbations are given by 
\begin{eqnarray}
\delta\left(\vk\right)&=& \delta_1\left(\vk\right)+
\int \frac{d^3\vq}{\left(2 \pi\right)^3}\delta_1\left(\vq\right)
\delta_1\left(\vk-\vq\right) \FtwoS\left(\vq,\vk-\vq\right)
\\ \nonumber & &+ \int \frac{d^3\vq_1}{\left(2 \pi\right)^3} 
\frac{d^3\vq_2}{\left(2 \pi\right)^3}
\delta_1\left(\vq_1\right)
\delta_1\left(\vq_2\right)
\delta_1\left(\vk-\vq_1-\vq_2\right) 
\FthreeS\left(\vq_1,\vq_2,\vk-\vq_1-\vq_2\right)
\label{eqgravPTbasic}
\end{eqnarray}
with
\begin{equation}
\FtwoS(\vk_1,\vk_2)=
\frac{5}{7}+\frac{1}{2}\frac{\vk_1\cdot\vk_2}{k_1 k_2}
\left(\frac{k_1}{k_2}+\frac{k_2}{k_1}\right)+\frac{2}{7}
\left(\frac{\vk_1\cdot\vk_2}{k_1 k_2}\right)^2
\end{equation}
and 
\begin{eqnarray}
F^{\left(3\right)}\left(\vq_1,\vq_2,\vq_3\right)&=&
\frac{1}{18}\left[
2~ \beta\left(\vq_1,\vq_2+\vq_3\right)~
G^{\left(2\right)}\left(\vq_2,\vq_3\right)+7~
\alpha\left(\vq_1,\vq_2+\vq_3\right)~
F^{\left(2\right)}\left(\vq_2,\vq_3\right) \right. \\ \nonumber &+& \left.
\left[2~\beta\left(\vq_1+\vq_2,\vq_3\right)~
+7~\alpha\left(\vq_1+\vq_2,\vq_3\right)\right]~
G^{\left(2\right)}\left(\vq_1,\vq_2\right) \right]~.
\end{eqnarray}
Note that this $F^{(3)}$ is un-symmetrized, while Eq. (\ref{eqgravPTbasic}) 
requires a symmetrized version (which we always indicate with a subscript
$S$). To symmetrize, average over all possible 
positionings of the arguments. See below for definitions of 
the component functions.

The velocity divergence $\theta$ is given by a similar expansion with
the kernels $F^{(N)}$ replaced by the following kernels $G^{(N)}$:
\begin{equation}
G_S^{(2)}(\vk_1,\vk_2)=
\frac{3}{7}+\frac{1}{2}\frac{\vk_1\cdot\vk_2}{k_1 k_2}
\left(\frac{k_1}{k_2}+\frac{k_2}{k_1}\right)+\frac{4}{7}
\left(\frac{\vk_1\cdot\vk_2}{k_1 k_2}\right)^2~, 
\end{equation}
and
\begin{eqnarray}
G^{\left(3\right)}\left(\vq_1,\vq_2,\vq_3\right)&=&
\frac{1}{18}\left[
6~\beta\left(\vq_1,\vq_2+\vq_3\right)~
G^{\left(2\right)}\left(\vq_2,\vq_3\right)+
3~\alpha\left(\vq_1,\vq_2+\vq_3\right)~
F^{\left(2\right)}\left(\vq_2,\vq_3\right) \right. \\ \nonumber &+& \left.
\left[6~\beta\left(\vq_1+\vq_2,\vq_3\right)~
+3~\alpha\left(\vq_1+\vq_2,\vq_3\right)\right]~
G^{\left(2\right)}\left(\vq_1,\vq_2\right) \right]~.
\end{eqnarray}
Again, note that this is un-symmetrized.

To represent the difference between $\theta$ and $\delta$,
we define $D^{\left(N\right)}=G^{\left(N\right)}- F^{\left(N\right)}$,
\begin{equation}
D_S^{(2)}(\vk_1,\vk_2)=
\frac{2}{7}\left[\left(\frac{\vk_1\cdot\vk_2}{k_1 k_2}\right)^2-1\right]
=\frac{2}{7}\left[\ssqfac\left(\vk_1,\vk_2\right)-
\frac{2}{3}\right]~,
\label{eqDFrelation}
\end{equation}
where we have defined $\ssqfac$ to represent Fourier space products 
of the operator $\gamma_{ij}$, 
\begin{equation}
\gamma_{ij}\left(\vq\right)\gamma_{ji}\left(\vk\right)=
\ssqfac\left(\vq,\vk\right)=
\frac{\left(\vq\cdot\vk\right)^2}
{k^2 q^2}-\frac{1}{3}~.
\end{equation}
Note that
\begin{equation}
\int_{-1}^{1}d \mu~ \ssqfac\left(\vk,\vq\right)=0
\end{equation}
where $\mu=\vk\cdot \vq/k q$.

The un-symmetrized 2nd order kernels (appearing in the 3rd order kernels)
are
\begin{equation}
F^{\left(2\right)}\left(\vq_1,\vq_2\right)=\frac{1}{7}\left[5~
\alpha\left(\vq_1,\vq_2\right)+2~ \beta\left(\vq_1,\vq_2\right)\right]
\end{equation}
and
\begin{equation}
G^{\left(2\right)}\left(\vq_1,\vq_2\right)=\frac{1}{7}\left[3~
\alpha\left(\vq_1,\vq_2\right)+4~ \beta\left(\vq_1,\vq_2\right)\right]~,
\end{equation}
where, finally,
\begin{equation}
\alpha\left(\vq,\vk\right)=\frac{\left(\vq+\vk\right)\cdot\vq}{q^2}
\end{equation}
and
\begin{equation}
\beta\left(\vq,\vk\right)=\frac{\left|\vq+\vk\right|^2\vq\cdot\vk}{2 q^2 k^2}
~.
\end{equation}

\bibliography{cosmo,cosmo_preprints}

\end{document}